\documentclass[12pt]{iopart}
\usepackage{amsbsy,amssymb}
\usepackage{iopams}
\usepackage{epsf}

\begin{document}

\newcommand{\bea}{\begin{eqnarray}}
\newcommand{\eea}{\end{eqnarray}}
\newcommand{\LCO}{La$_2$CuO$_4$}
\newcommand{\CuO}{CuO$_2$}
\newcommand{\LSCO}{La$_{2-x}$Sr$_x$CuO$_4$}
\newcommand{\ri}{\rm {i}}
\newcommand{\rv}{\rm {v}}
\newcommand{\re}{\rm {e}}
\newcommand{\sech}{\rm {sech}}
\newcommand{\bk}{\boldsymbol {k}}
\newcommand{\bq}{\boldsymbol {q}}


\title[Magnetic susceptibility of~\LCO]{Magnetic susceptibility of the body-centred orthorhombic~\LCO~system}

\author{Kyrylo V Tabunshchyk\footnote{Present address: NINT, University of Alberta, Edmonton AB Canada; Permanent address: Institute for Condensed Matter Physics,
               Lviv, Ukraine} and R J Gooding}
\address{Department of Physics, Queen's University,
               Kingston ON K7L 3N6 Canada}

\ead{gooding@physics.queensu.ca}

\begin{abstract}
A model Hamiltonian representing the Cu spins in La$_2$CuO$_4$ in its low-temperature body-centred orthorhombic
phase, that includes both spin-orbit generated Dzyaloshinskii-Moriya interactions and interplanar exchange,
is examined within the RPA utilizing a Tyablikov decoupling of various high-order Green's functions. The
magnetic susceptibility is evaluated as a function of temperature and the parameters quantifying these interactions,
and compared to recently obtained experimental data of Lavrov, Ando and collaborators.
An effective Hamiltonian corresponding to a simple tetragonal structure is shown to
reproduce both the magnon spectra and the susceptibility of the more complicated
body-centred orthorhombic model.
\end{abstract}




\section{Introduction}

Experimental studies of the magnetic and electronic properties of the cuprates continue to
produce new and unexpected results that spur on theorists in their attempts to understand
and describe the underlying orderings and excitations that may be involved in the pairing leading
to high-temperature superconductivity.
 One experiment, that was the motivation for the work that we present in this paper,
concerned the (zero-field) magnetic susceptibility of undoped La$_2$CuO$_4$ --
it was found \cite{Lavrov} that the magnetic response of
undoped ~\LCO~was highly anisotropic, and that this anisotropy persisted well above the
N\'eel ordering temperature.
 Further, they found that this anisotropy persisted in the weakly doped state.

 The importance of this result can be recognized if one notes the ongoing efforts of various
researchers in understanding the origin and nature of so-called stripe correlations that are
found in some cuprates (for a recent review of this problem, see reference \cite{jtran05}).
That is, if the undoped state has a highly anisotropic magnetic susceptibility, can it really
be that surprising that ``spin stripes" are also present when the system is doped, and if not, what role does the anisotropic magnetic response play in the formation of
stripe-like structures?

 Previously, we have examined \cite{Tabunshchyk} the origin of this magnetic anisotropy by considering
a single CuO$_2$ plane utilizing a magnetic Hamiltonian that contains spin-orbit generated
Dzyaloshinskii-Moriya (DM) interactions \cite{Dzyaloshinskii,Moriya}, and near-neighbour superexchange.
 If one includes both the symmetric and anti-symmetric DM interactions one finds that a true phase transition
(at a non-zero temperature) occurs to an antiferromagnetic (AFM) state, wherein the AFM moment lies in
the plane, with a weak parasitic ferromagnetic moment generated by a small canting of the moments out
of the plane.
 Within mean-field theory, linear spin-wave theory, and within the RPA utilizing the Tyablikov decoupling
scheme, we determined the magnetic susceptibility, and found (i) that it was indeed highly anisotropic,
even when the DM interactions were small compared to near-neighbour intraplanar superexchange,
and (ii) quantum fluctuations produced a substantial modification of the susceptibility as one used a
more and more ``sophisticated" theory \cite{Tabunshchyk}.
 Other potentially important terms (\emph{e.g.}, cyclic ring exchange \cite{Katanin}) that could have been
included in that paper are discussed at the end of this paper.

In this report we focus on an augmented model that now includes the third dimension and the full body-centred
orthorhombic structure of ~\LCO.
 Our motivations for doing so are as the following.
 I -- Although one can produce a true phase transition within a model that accounts for only a single plane,
the interplanar exchange interactions can also produce a phase transition in approximately the same temperature
range.
 That is, some researchers have suggested that both the (unfrustrated) interplanar exchange and the DM
interactions are of comparable strength, and thus there is no good reason to exclude either of these terms in
our model Hamiltonian (see \cite{Johnston} and references therein).
 II -- As we discuss below, there are two different ``near''-neighbour interplanar exchange constants, and
these are different in different directions. Thus, this difference will be a source of magnetic anisotropy,
and it is necessary to determine the extent of this anisotropy through a calculation that includes both the DM
interactions and the interplanar exchange.
 III -- As mentioned above, our previous work noted the strong effect of quantum fluctuations
in a two-dimensional model.
 Since one expects such effects to be larger the lower the dimensionality of the system, it is
possible that this behaviour is reduced in a full three-dimensional model.
 In this paper, again with the RPA utilizing the Tyablikov decoupling scheme, we have completed the requisite
calculations for this more complicated but also more realistic magnetic Hamiltonian.

Our paper is organized as follows. In the next section we summarize the formalism necessary to analyze
this problem; although somewhat similar formalism is presented our previous paper \cite{Tabunshchyk},
when going from 2D to 3D the analysis is much more complicated, and it is thus necessary to
present the required equations that must be solved. (Some aspects of the calculations
have been put into various appendices.) In the subsequent section we present the results of a detailed
and exhaustive numerical study of the resulting formalism for reasonable parameter values. Then we
suggest a simpler model Hamiltonian, one for a simple tetragonal structure which avoids the frustrated
interplanar AFM interactions of the original body-centred orthorhombic structure. Finally we conclude
the paper by discussing the key results that we have obtained, and then provide a comparison between
the predictions of our theory and the experiments of Lavrov, Ando and co-workers \cite{Lavrov}.

\section{Model and Methods}
\label{sec:Model}

\subsection{Model Hamiltonian}
\label{subsec:Model_IR}

\begin{figure}[h]
\centerline{\epsfxsize 1.\textwidth\epsfbox{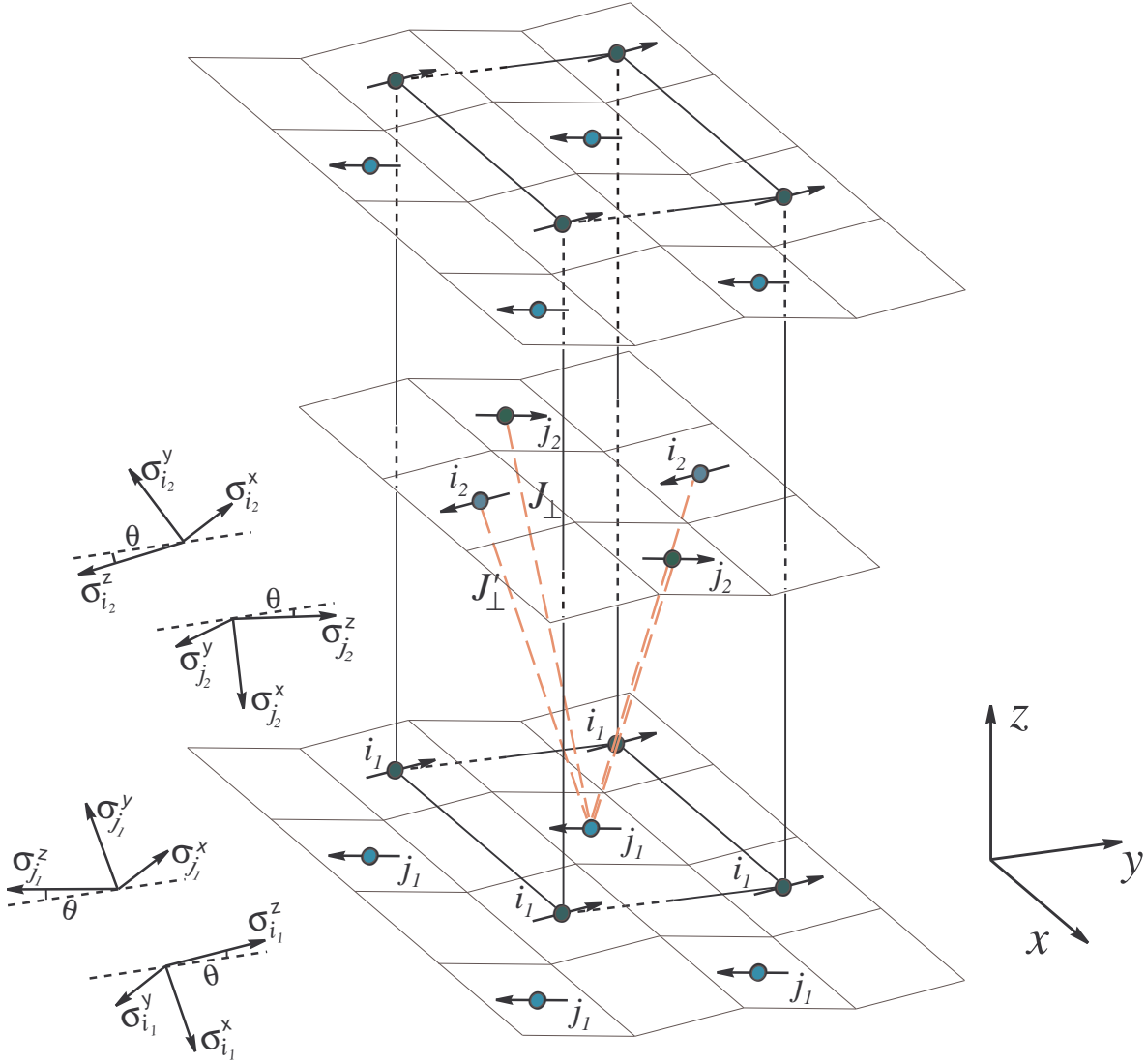}}
\caption{\label{fig:lattice}(Colour online) Magnetic structure of \LCO~crystal.
         Sites having different spin orientations are labelled by indices
         $i_1$, $j_1$ in the plane with the WF moment in the positive $z$ direction,
         and $i_2$, $j_2$ in the plane with the WF moment in the negative $z$ direction.
         For each set of sites the $\sigma$ spin coordinate system within the characteristic
         representation (CR) is shown.
         The thin net is shown only to simplify the visualization of the canting
         in the spin structure.}
\end{figure}

\begin{figure}[h]
\centerline{
\epsfxsize 0.8\textwidth\epsfbox{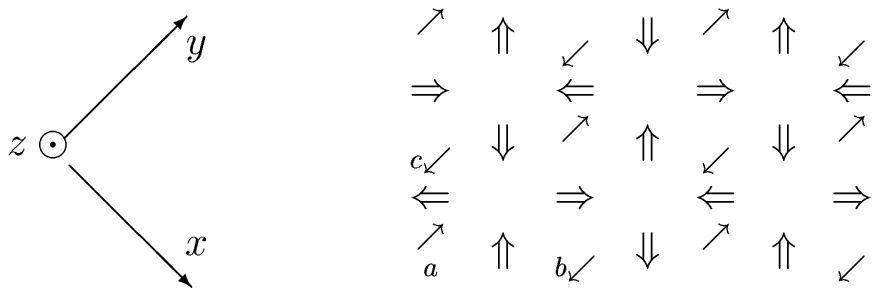}}
\hspace*{4cm}(a)\hspace*{4cm}(b)
\caption{\label{fig:vectors} (a) Coordinates in the initial representation.
                             (b) Thin arrows --- the Cu spins, and open arrows --- the DM vectors.}
\end{figure}

We describe the magnetic structure of the \LCO~crystal in the low-temperature
orthorhombic (LTO) phase by using an effective spin-$\frac12$ Hamiltonian
for the Cu$^{2+}$ magnetic ions of the \CuO~planes defined by
\numparts
\bea
\label{eq:H_DMa}
\fl   H&=&
       J\sum_{\langle i_1,j_1\rangle}{\bf S}_{i_1}\cdot{\bf S}_{j_1}
       +\sum_{\langle i_1,j_1\rangle}{\bf D}_{i_1j_1}\cdot({\bf S}_{i_1}\times{\bf S}_{j_1})
       +\sum_{\langle i_1,j_1\rangle}{\bf S}_{i_1}\cdot
                 \tilde{\Gamma}_{i_1j_1}\cdot{\bf S}_{j_1}\\
\label{eq:H_DMb}
\fl  &+&J\sum_{\langle i_2,j_2\rangle}{\bf S}_{i_2}\cdot{\bf S}_{j_2}
       +\sum_{\langle i_2,j_2\rangle}{\bf D}_{i_2j_2}\cdot({\bf S}_{i_2}\times{\bf S}_{j_2})
       +\sum_{\langle i_2,j_2\rangle}{\bf S}_{i_2}\cdot
                 \tilde{\Gamma}_{i_2j_2}\cdot{\bf S}_{j_2}\\
\label{eq:H_DMc}
\fl   &+&J_{\perp}\bigg\{\!
                          \sum_{\langle i_1,i_2\rangle}\!{\bf S}_{i_1}{\cdot}{\bf S}_{i_2}
                         {+}\sum_{\langle j_1,j_2\rangle}\!{\bf S}_{j_1}{\cdot}{\bf S}_{j_2}
                   \!\bigg\}
         +J'_{\perp}\bigg\{\!
                           \sum_{\langle i_1,j_2\rangle}\!{\bf S}_{i_1}{\cdot}{\bf S}_{j_2}
                          {+}\sum_{\langle j_1,i_2\rangle}\!{\bf S}_{j_1}{\cdot}{\bf S}_{i_2}
                   \!\bigg\}~~.
\eea
\endnumparts
 In this equation ${\bf S}_{i}$ denotes a spin at site $i$, and sites labelled as $i_1$ and $j_1$
are in the ``first" plane while $i_2$ and $j_2$ are in the ``second" (neighbouring) plane;
the notation $\langle i_\alpha,j_\beta\rangle$ refer to near-neighbour sites.
 This Hamiltonian is written within the $xyz$ orthorhombic coordinate system
shown in figure~\ref{fig:lattice} (see right-hand side) and in figure~\ref{fig:vectors}(a),
in what we refer to as the ``initial representation'' in the LTO phase.

The various terms in the magnetic Hamiltonian given in equation (1) correspond to the following
interactions.
 As was mentioned in the introduction, the orthorhombic distortion in the \LCO~crystal,
together with the spin-orbit coupling, lead to the antisymmetric
Dzyaloshinskii-Moriya ($\bf D$~term) and the symmetric
pseudo-dipolar ($\tilde{\Gamma}$~term) interactions
within the each \CuO~plane \cite{Coffey,Aharony,Koshibae}.
 These interactions together with su\-per\-ex\-chan\-ge one ($J$) can give
rise to an ordered phase within a \emph{single} \CuO~plane at some
nonzero temperature \cite{Tabunshchyk}.
 In this long-range ordered state Cu spins are aligned antiferromagnetically
in the $y$-direction, with a small canting out of the plane.
 Therefore, each \CuO~plane in a \LCO~crystal exhibits a net ferromagnetic
moment, so-called weak ferromagnetism (WF) in the direction
parallel to the $c$-axis of the {\it{Bmab}} space group ($z$-axis
in the initial coordinates).
 Due to the weak antiferromagnetic coupling between the planes, the net ferromagnetic
moments of adjacent \CuO~planes are antiferromagnetically aligned and the system possesses
no net moment.

 Each Cu spin has four sites above and below it in neighbouring planes.
 If all of these distances were equal the system would be frustrated because the ordering in one
plane would not lift the degeneracy that would result in adjacent planes.
 However, in the LTO phase these distances are not all equal, and thus the interplanar coupling
between nearest-neighbour spins depends on which pair of neighbouring sites are considered.
 That is, due to the small orthorhombic distortion (relative to the high-temperature
body-centred tetragonal phase) some near neighbour sites are closer together than other pairs
(which are, technically, next-near-neighbour sites).
 In what follows we refer to the sites shown in figure 1, which allows for these ideas and the
interplanar terms in equation~(\ref{eq:H_DMc}) to be made clear.
 The distance between $j_1$ and $j_2$ sites is slightly less than the distance between
$j_1$ and $i_2$, and thus the superexchange couplings are different, and in this paper
we thus specify that neighbouring spins in the $x-z$ plane ($J_{\perp}$ ) have a larger
superexchange than do neighbouring spins in the $y-z$ plane ($J'_{\perp}$): $|J_{\perp}|>|J'_{\perp}|$
(see, for example, the discussion in reference \cite{Xue}).
 As discussed in the introduction, and quantified in the next section,
this difference immediately leads to an enhanced anisotropy of the magnetic
susceptibility.

  We schematically illustrate the magnetic structure of the \LCO~crystal within
the ordered state ($T<T_N$) in figure~\ref{fig:lattice}, where arrows represent
the Cu spin structure; this ordered state is quadripartite, as we now explain.
 In our notations we label sites in a plane with the spin canting up as $i_1$ and $j_1$,
and correspondingly the sites of the nearest-neighbour planes with the spin canting down
are labelled by indices $i_2$ and $j_2$.
 In each plane sites with label $i$ differ from the sites $j$ by the spin orientation
within the antiferromagnetic order.
 Clearly, the magnetic structure of the \LCO~crystal in the ground state can be
represented by four different sublattices with different spin orientations, and
in our calculations we will follow the notation that $i_1$-sites belong to
sublattice~1, $j_1$-sites to sublattice~2, $i_2$-sites to sublattice~3,
and $j_2$-sites to sublattice~4.
 The interaction of the spins from the sublattice 1 and 2 with the
nearest-neighbour spins from sublattice 3 and 4, respectively, are described by term $J_{\perp}$,
and interaction with the spins from 4 and 3, respectively, are described by term $J'_{\perp}$.
 Each magnetic ion interacts with the four nearest neighbour sites within its plane and with
eight ions (four above and four below) from neighbouring planes.

 To summarize, the above presented magnetic Hamiltonian describes the magnetic interactions
within the each \CuO~plane in its first and second parts (equations~(\ref{eq:H_DMa},~\ref{eq:H_DMb})),
while the third part (equation~(\ref{eq:H_DMc})) takes
into account the weak interplanar superexchange couplings.

 The structure of the Dzyaloshinskii-Moriya (DM) and the pseudo-dipolar interactions
for the LTO phase are given by
\bea
\label{eq:DM}
    {\bf D}_{ab}=\frac d{\sqrt{2}}(-1,1,0),\qquad
    {\bf D}_{ac}=\frac d{\sqrt{2}}(-1,-1,0),
\eea
and
\bea
\label{eq:Gamma}
\tilde{\Gamma}_{ab}=
  \left( \begin{array}{ccc}
     \Gamma_1 & \Gamma_2 & 0 \\
     \Gamma_2 & \Gamma_1 & 0 \\
     0 & 0 & \Gamma_3
         \end{array} \right),\qquad
\tilde{\Gamma}_{ac}=
  \left( \begin{array}{ccc}
     \Gamma_1 & -\Gamma_2 & 0 \\
     -\Gamma_2 & \Gamma_1 & 0 \\
     0 & 0 & \Gamma_3
         \end{array} \right),
\eea within the initial coordinate system \cite{Tabunshchyk}.
 The DM vector given in equation~(\ref{eq:DM}) alternates in sign on successive bonds in the $a-b$
and in the $a-c$ direction of each plane, as is represented schematically by the double arrows in
figure~\ref{fig:vectors}(b).
 Thin arrows in this figure describe the in-plane antiferromagnetic order of
the Cu spins, and are canted up/down from the in-plane order by a small angle.
 In the classical ground state of the LTO phase the absolute value of the canting angles are
equal on all sites and are given by the expression
\bea
\label{eq:canting-angle}
\theta = \frac 12\tan^{-1}\Big\{\frac{d/\sqrt{2}}{J+\frac 12(\Gamma_1+\Gamma_3)-\frac 12J'_{\perp}}\Big\}.
\eea

 Following the scheme described in our earlier work \cite{Tabunshchyk}, we perform rotations of the spin
coordinate system in such a way that the new quantization axis ($\sigma^z$) is in the direction
of a classical moment characterizing the ground state.
 Hereinafter we will call such a representation as the ``characteristic representation'' (CR).
 Since four different types of spin orientations are present in the magnetic structure
of the \LCO~crystal, we introduce four different spin coordinate systems (see the
left-hand side of figure~\ref{fig:lattice})
given by the transformations equations~(\ref{eq:1_rot1}-\ref{eq:1_rot4}) in Appendix~A.
 Thus, each sublattice consists its own spin coordinate system.
 The model Hamiltonian in terms of these spin operators $\sigma$ in the CR is given by
\bea
\nonumber
\fl H_{\rm CR} &=& \sum_{\langle i_1,j_1\rangle_{ab}}
    \left\{ A(\sigma^+_{i_1}\sigma^-_{j_1}+\sigma^-_{i_1}\sigma^+_{j_1})
    -B^*\sigma^+_{i_1}\sigma^+_{j_1}
    -B\sigma^-_{i_1}\sigma^-_{j_1}
    -J_2\sigma^z_{i_1}\sigma^z_{j_1}
    \right\}\\
\nonumber
\fl &+&  \sum_{\langle i_1,j_1\rangle_{ac}}
    \left\{ A(\sigma^+_{i_1}\sigma^-_{j_1}+\sigma^-_{i_1}\sigma^+_{j_1})
    +B\sigma^+_{i_1}\sigma^+_{j_1}
    +B^*\sigma^-_{i_1}\sigma^-_{j_1}
    -J_2\sigma^z_{i_1}\sigma^z_{j_1}
    \right\}\\
\label{eq:H_CR}
\fl &+&\sum_{\langle i_2,j_2\rangle_{ab}}
    \left\{ A(\sigma^+_{i_2}\sigma^-_{j_2}+\sigma^-_{i_2}\sigma^+_{j_2})
    +B\sigma^+_{i_2}\sigma^+_{j_2}
    +B^*\sigma^-_{i_2}\sigma^-_{j_2}
    -J_2\sigma^z_{i_2}\sigma^z_{j_2}
    \right\}\\
\nonumber
\fl &+&  \sum_{\langle i_2,j_2\rangle_{ac}}
    \left\{ A(\sigma^+_{i_2}\sigma^-_{j_2}+\sigma^-_{i_2}\sigma^+_{j_2})
    -B^*\sigma^+_{i_2}\sigma^+_{j_2}
    -B\sigma^-_{i_2}\sigma^-_{j_2}
    -J_2\sigma^z_{i_2}\sigma^z_{j_2}
    \right\}\\
\nonumber
\fl &+& \!\!\sum_{\langle i_1,j_2\rangle}
    \left\{\frac14({J'_{\perp}}{+}J_p)(\sigma^+_{i_1}\sigma^-_{j_2}+\sigma^-_{i_1}\sigma^+_{j_2})
     +{\ri}\frac14({J'_{\perp}}{-}J_p)(\sigma^+_{i_1}\sigma^+_{j_2}-\sigma^-_{i_1}\sigma^-_{j_2})
       +J_p\sigma^z_{i_1}\sigma^z_{j_2}
\right\}\\
\nonumber
\fl &+& \!\!\sum_{\langle j_1,i_2\rangle}
    \left\{\frac14({J'_{\perp}}{+}J_p)(\sigma^+_{j_1}\sigma^-_{i_2}+\sigma^-_{j_1}\sigma^+_{i_2})
     +{\ri}\frac14({J'_{\perp}}{-}J_p)(\sigma^+_{j_1}\sigma^+_{i_2}-\sigma^-_{j_1}\sigma^-_{i_2})
       +J_p\sigma^z_{j_1}\sigma^z_{i_2}
\right\}\\
\nonumber
\fl &+& J_{\perp}\!\!\sum_{\langle i_1,i_2\rangle}
       \left\{\frac{\ri}{2}(\sigma^+_{i_1}\sigma^+_{i_2}{-}\sigma^-_{i_1}\sigma^-_{i_2}){-}
       \sigma^z_{i_1}\sigma^z_{i_2}\right\}
     + J_{\perp}\!\!\sum_{\langle j_1,j_2\rangle}
       \left\{\frac{\ri}{2}(\sigma^+_{j_1}\sigma^+_{j_2}{-}\sigma^-_{j_1}\sigma^-_{j_2}){-}
       \sigma^z_{j_1}\sigma^z_{j_2}\right\},
\eea
where we have used the following definitions:
\bea
\label{eq:AB}
\fl   && A = \frac{J_1-J_3}4,\qquad
      B = \frac{J_4}2+{\rm i} \frac{J_1+J_3}4,\\
\label{eq:J1}
\fl   &&J_1=J+\Gamma_1, \qquad  J_p= J'_{\perp}\cos2\theta,\\
\label{eq:J2}
\fl   && J_2=\hphantom{-}
             \frac {\Gamma_1{-}\Gamma_3}2+\left(J+\frac {\Gamma_1{+}\Gamma_3}2\right)\cos2\theta
                                         +\frac{d}{\sqrt{2}}\sin2\theta,\\
\label{eq:J3}
\fl   && J_3=-\frac {\Gamma_1{-}\Gamma_3}2+\left(J+\frac {\Gamma_1{+}\Gamma_3}2\right)\cos2\theta
                                         +\frac{d}{\sqrt{2}}\sin2\theta,\\
\label{eq:J4}
\fl   && J_4=-\Gamma_2\sin\theta+\frac d{\sqrt{2}}\cos\theta.
\eea
The subscripts $\langle i,j\rangle_{ab}$ and $\langle i,j\rangle_{ac}$
in the summations of equation~(\ref{eq:H_CR}) imply the nearest neighbours in the
$ab$ and $ac$ directions, respectively, as shown in figure~\ref{fig:vectors}(b).

\subsection{Mean field analysis}

 In this subsection, we present the results of the mean field approximation (MFA) for the above
Hamiltonian by following the standard decoupling. That is, in the MFA in equation (1) we use
\bea
\label{eq:MFA_decoupling}
\sigma_i^a\sigma_j^b \rightarrow
\langle \sigma_i^a\rangle ~ \sigma_j^b ~+~
\sigma_i^a ~ \langle \sigma_j^b\rangle ~-~
\langle \sigma_i^a\rangle ~ \langle \sigma_j^b\rangle,
\eea
where $a$ and $b$ can be equal to any of $x,y,z$.
 Then, the equation for the order parameter, to be denoted by $\eta$, within the MFA
reads as
\bea
\label{eq:sigma_MFA}
\eta\equiv\langle\sigma^z\rangle = \frac 12 \tanh \left\{\frac{\beta}2
                              {\cal Z}[J_2+J_{\perp}-J_p]\langle\sigma^z\rangle\right\},
\eea
where ${\cal Z}=4$ is the in-plane coordination number, and $\beta = 1/T$.
 From this equation the N\'eel temperature at which $\eta$ vanishes can be
written immediately as
\bea
\label{eq:T_N^MFA}
 T^{MFA}_{N}  = J_2+J_{\perp}-J_p~~.
\eea
 By applying a magnetic field sequentially in the $x$, $y$, and $z$ directions of each
coordinate systems within the CR we can find the transverse and
longitudinal components of susceptibility within all four sublattices.
 Using the relation between the components of susceptibility in the initial
and characteristic representations given in equations~(\ref{eq:CRtoINx}-\ref{eq:CRtoINz}),
we obtain the final result for the zero-field uniform susceptibility within the MFA below
the ordering temperature ($T<T^{MFA}_N$)
\bea
\label{eq:MFA_SxSx}
\fl\chi^{x~ MFA}&=&\frac 14 \frac 1{J_1 {+} J_2 {+} 2J_{\perp} {+} (J'_{\perp} {-} J_p)},\\
\fl\chi^{y~ MFA}&=&\frac 14 \frac {\sin^2(\theta)}
                                   {J_2 {-} J_3 {+} 2J_{\perp}{-} 2J_p}
           +\frac{\cos^2\theta}{4}
       \frac{{\sech}^2\left\{\frac{\beta}2{\cal Z}\eta J_{mfa}\right\}}
            {T+[J_2{+}J_{\perp}{+}J_p]\:
             {\sech}^2\left\{\frac{\beta}2{\cal Z}\eta J_{mfa}\right\}},
\label{eq:MFA_SySy}\\
\fl\chi^{z~ MFA}&=&\frac 14 \frac {\cos^2(\theta)}
                                   {J_2 {+} J_3 {+} 2J_{\perp}}
           +\frac{\sin^2\theta}{4}
       \frac{{\sech}^2\left\{\frac{\beta}2{\cal Z}\eta J_{mfa}\right\}}
            {T-[J_2{-}J_{\perp}{+}J_p]\:
             {\sech}^2\left\{\frac{\beta}2{\cal Z}\eta J_{mfa}\right\}},
\label{eq:MFA_SzSz}
\eea
where we define
\bea
\label{J_mfa}
J_{mfa}= J_2+J_{\perp}-J_p~~,
\eea
and the equation for the order parameter $\eta$ given by equation~(\ref{eq:sigma_MFA}).
 We used the ``\emph{mfa}'' subscript in equation~(\ref{J_mfa}) because this combination determines
the effective interaction, and thus the N\'eel temperature within the mean field theory
(see equation~(\ref{eq:sigma_MFA})).

 The final results for the components of the susceptibility in the initial representation
for high temperatures, that is above the ordering temperature ($T>T^{MFA}_N$), are
\bea
\label{eq:MFA_SxSx_para}
\chi^{x~ MFA}&=&\frac 14 \frac 1{T + J_1 + J_{\perp} + J'_{\perp}},\\
\label{eq:MFA_SySy_para}
\chi^{y~MFA}&=&\frac 14 \frac {\sin^2(\theta)}
                                {T - J_3 + J_{\perp} - J_p}
           +\frac 14 \frac{\cos^2(\theta)}{T+J_2+J_{\perp}+J_p},\\
\label{eq:MFA_SzSz_para}
\chi^{z~MFA}&=&\frac 14 \frac {\cos^2(\theta)}
                                {T + J_3 + J_{\perp} + J_p}
           +\frac 14 \frac{\sin^2(\theta)}{T-J_2+J_{\perp}-J_p}.
\eea
 In the limit $T\to T^{MFA}_{\rm N}$ we obtain that the
$x$ component of the susceptibility is continuous at the
transition and is given by equation~(\ref{eq:MFA_SxSx}).
 The $y$ component of the susceptibility at the transition
temperature reads
\bea
  \chi^{y~ MFA}\bigg|_{T\to T^{MFA}_{\rm N}}
              =\frac 14 \frac {\sin^2(\theta)}{J_2 - J_3 + 2J_{\perp} - 2J_p}
              +\frac{\cos^2\theta}{8}\frac{1}{J_{\perp} + J_2}.
\label{eq:MFA_SySy_Tn}
\eea
 Note that with respect to the pure 2D case \cite{Tabunshchyk} the $z$ component of
the susceptibility does not diverge at the N\'eel point and also is
continuous at the transition
\bea
  \chi^{z~ MFA}\bigg|_{T\to T^{MFA}_{\rm N}}
              =\frac 14 \frac {\cos^2(\theta)}{J_2 + J_3 + 2J_{\perp}}
              +\frac 18 \frac{\sin^2(\theta)}{J_{\perp} - J_p}.
\label{eq:MFA_SzSz_Tn}
\eea

\subsection{Random phase approximation}
 In this part of the paper we use the technique of the double-time temperature dependent
Green's functions within the framework of the random-phase approximation (RPA).
 In the imaginary-time formalism, the temperature dependent Green's function
and the corresponding equation of motion for two Bose-type operators reads
\bea
\label{eq:defG}
\fl G_{AB}(\tau) = \langle T_\tau A(\tau)B(0)\rangle,\quad
  \frac{{\rm d}G_{AB}(\tau)}{{\rm d}\tau}=\delta(\tau)\langle[A,B]\rangle
  +\langle T_\tau[H(\tau),A(\tau)]B(0)\rangle,
\eea
where $A(\tau)={\rm e}^{H\tau}A{\rm e}^{-H\tau}$ is the operator in the Heisenberg
representation for imaginary time argument $\tau$, and $T_\tau$ is the time-ordering
operator.

 By using the method proposed originally by Liu \cite{Liu}, we employ the perturbed
Hamiltonian
\bea
\label{eq:H_pert}
H^f_1 = H_{\rm CR} - f \sum_{i'_1}\sigma^z_{i'_1},\qquad i'_1 \in \mbox{ sublattice 1}
\eea
to find the longitudinal components of the susceptibility in the CR.
 In the equation~(\ref{eq:H_pert}) $f$ is a small fictitious field applied to the
spins of {\emph {sublattice 1 only}}.
 In this paper we are studying the zero-field uniform magnetic susceptibility,
therefore we restrict $f$ to be constant and static.

 The Green's functions to be used in present calculations are
\begin{eqnarray}
 \nonumber
\fl   &&G^f_{i_1j_1}(\tau)=\langle T_{\tau}\sigma^+_{i_1}(\tau)\sigma^-_{j_1}(0)\rangle^f,\quad
    G^{f-}_{i_1j_1}(\tau)=\langle T_{\tau}\sigma^-_{i_1}(\tau)\sigma^-_{j_1}(0)\rangle^f,\\
\label{eq:GGG}
\fl  &&G^f_{j'_1j_1}(\tau)=\langle T_{\tau}\sigma^+_{j'_1}(\tau)\sigma^-_{j_1}(0)\rangle^f,\quad
   G^{f-}_{j'_1j_1}(\tau)=\langle T_{\tau}\sigma^-_{j'_1}(\tau)\sigma^-_{j_1}(0)\rangle^f,\\
 \nonumber
\fl   &&G^f_{i_2j_1}(\tau)=\langle T_{\tau}\sigma^+_{i_2}(\tau)\sigma^-_{j_1}(0)\rangle^f,\quad
    G^{f-}_{i_2j_1}(\tau)=\langle T_{\tau}\sigma^-_{i_2}(\tau)\sigma^-_{j_1}(0)\rangle^f,\\
 \nonumber
\fl  &&G^f_{j'_2j_1}(\tau)=\langle T_{\tau}\sigma^+_{j'_2}(\tau)\sigma^-_{j_1}(0)\rangle^f,\quad
   G^{f-}_{j'_2j_1}(\tau)=\langle T_{\tau}\sigma^-_{j'_2}(\tau)\sigma^-_{j_1}(0)\rangle^f,
\end{eqnarray}
where $\langle...\rangle^f$ means that all expectation values are taken with respect to the perturbed
Hamiltonian in equation~(\ref{eq:H_pert}).
 After an expansion in a power series of $f$ the Green's function, \emph{e.g.} $G^f_{i_1j_1}(\tau)$, reads
\bea
\label{eq:def_Gf}
 G^f_{i_1j_1}(\tau) = G^{(0)}_{i_1j_1}(\tau) + f G^{(1)}_{i_1j_1}(\tau) + O(f^2).
\eea
Since $G^{(0)}_{i_1j_1}(\tau)=G_{i_1j_1}(\tau)$, from now we drop the superscript and use
\bea
\label{eq:explanation1}
 G^f_{i_1j_1}(\tau) = G_{i_1j_1}(\tau) + f G^{(1)}_{i_1j_1}(\tau) + O(f^2).
\eea
Also, we introduce
\bea
\label{eq:explanation2}
 \langle \sigma^z_{i_1}(\tau)\rangle^f =
   \langle \sigma^z_{i_1}\rangle + f{\rm v}_{i_1}+O(f^2),
\eea
where, due to the translation periodicity
$\langle \sigma^z_{i_1}\rangle=\eta$, the order parameter at $f=0$.

 Now let us find the equation of motion for the Green's function
$G^f_{i_1j_1}(\tau)$. The equations for other functions can be found in the
same way.
 Starting from the equation~(\ref{eq:defG}) we can write
\begin{eqnarray}
\label{eq:eqom_Gf}
\fl \frac{{\rm d}G^f_{i_1j_1}(\tau)}{{\rm d}\tau}=2\delta(\tau)\delta_{i_1j_1}
  \langle \sigma^z_{i_1}\rangle^f
  +\langle T_\tau[H_{\rm CR}(\tau),\sigma^+_{i_1}(\tau)]\sigma^-_{j_1}(0)\rangle^f
  -fG^f_{i_1j_1}.
\end{eqnarray}
 In order to solve this equation of motion we are following the RPA scheme, and using
the so-called Tyablikov's decoupling \cite{Tyablikov} which is given by
\bea
\label{eq:Tyablikov_decoupling}
\fl\langle T_{\tau} \sigma^z_l(\tau)\sigma^+_{i_1}(\tau)\sigma^-_{j_1}(0) \rangle^f
 \to \langle \sigma^z_l(\tau)\rangle^f
 \langle T_{\tau} \sigma^+_{i_1}(\tau)\sigma^-_{j_1}(0) \rangle^f
 =\langle \sigma^z_l(\tau)\rangle^f G^f_{i_1j_1}(\tau).
\eea
After this decoupling is introduced, equation~(\ref{eq:eqom_Gf}) is found to be
\begin{eqnarray}
\label{eq:eqom_GF_2}
\fl \frac{{\rm d}G^f_{i_1j_1}(\tau)}{{\rm d}\tau}&=&
     2\delta(\tau)\delta_{i_1j_1}\langle \sigma^z_{i_1}\rangle^f -fG^f_{i_1j_1}(\tau)\\
\nonumber
\fl &&\hspace*{-.9cm}
    -\!\!\sum_{\delta_{ab}}\left\{\!2\langle \sigma^z_{i_1}(\tau)\rangle^f
    [AG^f_{(i_1{+}\delta)j_1}(\tau)-BG^{f-}_{(i_1{+}\delta)j_1}(\tau)]
    +J_2\langle \sigma^z_{i_1{+}\delta}(\tau)\rangle^fG^f_{i_1j_1}(\tau)\!\right\}\\
\nonumber
\fl &&\hspace*{-.9cm}-\!\!\sum_{\delta_{ac}}\left\{2\langle \sigma^z_{i_1}(\tau)\rangle^f
    [AG^f_{(i_1{+}\delta)j_1}(\tau){+}B^*G^{f-}_{(i_1{+}\delta)j_1}(\tau)]
    +J_2\langle \sigma^z_{i_1{+}\delta}(\tau)\rangle^fG^f_{i_1j_1}(\tau)\right\}\\
\nonumber
\fl &&\hspace*{-.9cm}-\!\!\sum_{\langle j'_2\rangle_{i_1}}\left\{2\langle \sigma^z_{i_1}(\tau)\rangle^f
    \bigg[ \frac{J'_{\perp}{+}J_p}{4}G^f_{j'_2j_1}(\tau)
         -{\ri}\frac{J'_{\perp}{-}J_p}{4}G^{f-}_{j'_2j_1}(\tau)\bigg]
    -J_p\langle \sigma^z_{j'_2}(\tau)\rangle^fG^f_{i_1j_1}(\tau)\!\right\}\\
\nonumber
\fl &&\hspace*{-.9cm}-\!\!\sum_{\langle i'_2\rangle_{i_1}}\left\{-{\ri}J_{\perp}\langle \sigma^z_{i_1}(\tau)\rangle^fG^{f-}_{i'_2j_1}(\tau)
                                  +J_{\perp}\langle \sigma^z_{i'_2}(\tau)\rangle^fG^{f}_{i_1j_1}(\tau)\!\right\},
\end{eqnarray}
where $\sum_{\delta_{ab}}$ refers to a summation over the nearest neighbours
of the sites $i_1$ in the $ab$ direction of the same \CuO~plane, and similarly for
$\sum_{\delta_{ac}}$ --- see figure~\ref{fig:vectors}(b).
 Thus, in equation~(\ref{eq:eqom_GF_2}) all sites $i_1+\delta$ belong to the sublattice 2.
 The notation $\sum_{\langle i'_2\rangle_{i_1}}$ means sum over all sites $i'_2$ from sublattice 3
which are nearest neighbours of sites $i_1$ that belong to the sublattice 1,
and similarly for $\sum_{\langle j'_2\rangle_{i_1}}$.

 Next, we perform the transformation into the momentum-frequency representation
for the Green's functions and the spin operators:
\bea
\label{eq:Fourier1}
\fl G^f_{i_1j_1}(\tau)&=&\frac 4{N\beta}\sum_{\bk, m}G^f_{12}(\bk,\omega_m)
  {\rm e}^{{\rm i}\bk\cdot({\bf R}_{i_1}-{\bf R}_{j_1})}{\rm e}^{-{\rm i}\omega_m\tau},\\
\label{eq:Fourier2}
\fl \langle\sigma^z_{i_1}(\tau)\rangle^f&=&
   \frac 1{\beta}\sum_{\bk, m}\langle \sigma^z_1(\bk,\omega_m)\rangle^f
   {\rm e}^{-{\rm i}\bk\cdot{\bf R}_{i_1}}{\rm e}^{-{\rm i}\omega_m\tau}
   = \sum_{\bk}\delta(\bk)[\eta + f{\rm v}_1]{\rm e}^{-{\rm i}\bk\cdot{\bf R}_{i_1}},
\eea
where the expansion of equation~(\ref{eq:explanation2}) and the linear response to the uniform
perturbation expressed by ${\rm v}_1(\bk)=\delta(\bk){\rm v}_1$ were taken into account.
In the transformation given by equations~(\ref{eq:Fourier1},~\ref{eq:Fourier2}), the sum over
$\bk$ runs over $\frac 14 N$ points of the first Brillouin zone, and
$\omega_n = 2\pi n/\beta$ for $n\in\mathbb{Z}$ are the Bose Matsubara frequencies.
 The equation of motion for the Green's function $G^f_{i_1j_1}(\tau)$ in the
momentum-frequency representation reads
\bea
\label{eq:eq3}
\fl -{\rm i}\omega_m G^f_{12}(\bk,\omega_m)&=&
     -2{\cal Z}A_{\bk}[\eta + f{\rm v}_1]G^f_{22}(\bk,\omega_m)
            +2{\cal Z}B_{\bk}[\eta + f{\rm v}_1]G^{f-}_{22}(\bk,\omega_m)\\
\nonumber
\fl  &&     -2{\cal Z}\:a_{\bk}[\eta + f{\rm v}_1]G^f_{42}(\bk,\omega_m)
       +2{\cal Z}\:{\ri}b_{\bk}[\eta + f{\rm v}_1]G^{f-}_{42}(\bk,\omega_m)\\
\nonumber
\fl  &&+2{\cal Z}\:{\ri}d_{\bk}[\eta + f{\rm v}_1]G^{f-}_{32}(\bk,\omega_m)
       -fG^f_{12}(\bk,\omega_m)\\
\nonumber
\fl && -{\cal Z}\bigg\{ J_2      [\eta + f{\rm v}_2]
                       +J_{\perp}[\eta + f{\rm v}_3]
                       -J_p      [\eta + f{\rm v}_4] \bigg\}
                                      G^f_{12}(\bk,\omega_m),
\eea
where, as before, ${\cal Z}$ is the in-plane coordination number, and we introduce
\bea
\label{eq:notationAB}
 && A_{\bk} = A\gamma_{\bk},\quad
   B_{\bk} = (\Re B) \gamma_{\bk}'+{\ri}(\Im B) \gamma_{\bk}, \\
\label{eq:notationabd}
 && a_{\bk} = \frac{J'_{\perp}+J_p}4\xi_{\bk}, \quad
    b_{\bk} = \frac{J'_{\perp}-J_p}4\xi_{\bk}, \quad
    d_{\bk} = \frac{J_{\perp}}2\xi'_{\bk},\\
 && \gamma_{\bk}=\frac 12(\cos k_x+\cos k_y),\quad
       \xi_{\bk}= \cos k_z\cos\bigg(\frac{k_x+k_y}2\bigg),\\
 && \gamma'_{\bk}=\frac 12(\cos k_x-\cos k_y),\quad
    \xi'_{\bk}= \cos k_z\cos\bigg(\frac{k_x-k_y}2\bigg).
\eea
 Now we can write down the final equations for the zero-order in $f$ Green's function
$G_{12}(\bk,\omega_m)$, and the first-order one $G^{(1)}_{12}(\bk,\omega_m)$
\bea
\fl  \frac {{\rm i}\omega_m}{2{\cal Z}\eta}G_{12}&=&
   \bigg\{\frac{J_2}2{+}\frac{J_{\perp}}2{-}\frac{J_p}2\bigg\}G_{12}
  +A_{\bk}G_{22}-B_{\bk}G^-_{22}
  +a_{\bk}G_{42}-{\ri}b_{\bk}G^-_{42}-{\ri}d_{\bk}G^-_{32},\\
\nonumber
\fl   \frac {{\rm i}\omega_m}{2{\cal Z}\eta}G^{(1)}_{12}&=& \frac{G_{12}}{2{\cal Z}\eta}+
                                          \bigg\{\frac{J_2}2 \frac{{\rv}_2}{\eta}
                                          {+}\frac{J_\perp}2 \frac{{\rv}_3}{\eta}
                                              {-}\frac{J_p}2 \frac{{\rv}_4}{\eta}\bigg\}G_{12}
+\frac{{\rv}_1}{\eta}\bigg(\frac{{\ri}\omega}{2{\cal Z}\eta}
                    -\bigg\{\frac{J_2}2{+}\frac{J_{\perp}}2{-}\frac{J_p}2\bigg\}
                     \bigg)G_{12}\\
\nonumber
\fl &+& \bigg\{\frac{J_2}2{+}\frac{J_{\perp}}2{-}\frac{J_p}2\bigg\}G^{(1)}_{12}
    +A_{\bk}G^{(1)}_{22}-B_{\bk}G^{(1)-}_{22}
    +a_{\bk}G^{(1)}_{42}-{\ri}b_{\bk}G^{(1)-}_{42}-{\ri}d_{\bk}G^{(1)-}_{32}\\
\fl &&
\eea
where in these equations we drop the wave vector and frequency dependencies
for the Green's functions; that is $G=G(\bk,\omega_m)$
and $G^{(1)}=G^{(1)}(\bk,\omega_m)$.

 In order to obtain a closed set of the equations for the zero and first order
Green's function we should use the above described scheme for the all other functions
in equation~(\ref{eq:GGG}), and
 the final system of equations for the zero and first-order Green's function are given
in the Appendix B in equations~(\ref{eq:system_g}-\ref{eq:coef_G1}).
 The structure of the system for the zero-order functions is identical with
the system of equations for the first-order ones, except for the free terms.
 Hence, the poles of the zero-order Green's functions (that determine the spectrum of
the spin-wave excitations) $G(\bk,\omega_m)$ are equal to the poles for the first-order
ones $G^{(1)}(\bk,\omega_m)$, and are found to be
\bea
\label{eq:varepsilon}
\fl&&\varepsilon_{1,\bk} = 2{\cal Z}\eta\omega_{1,\bk}=\sqrt{\alpha_{1,\bk}+\sqrt{\beta_{1,\bk}}}~~,\qquad
     \varepsilon_{2,\bk} = 2{\cal Z}\eta\omega_{2,\bk}=\sqrt{\alpha_{1,\bk}-\sqrt{\beta_{1,\bk}}}~~,\\
\nonumber
\fl&&\varepsilon_{3,\bk} = 2{\cal Z}\eta\omega_{3,\bk}=\sqrt{\alpha_{2,\bk}+\sqrt{\beta_{2,\bk}}}~~,\qquad
     \varepsilon_{4,\bk} = 2{\cal Z}\eta\omega_{4,\bk}=\sqrt{\alpha_{2,\bk}-\sqrt{\beta_{2,\bk}}}~~,\\
\label{eq:alpha}
\fl&&\alpha_{1,\bk} = a_{\bk}^2+(A_{\bk}{-}J_{mfa}/2)^2 -(b_{\bk}{-}d_{\bk})^2-|B_{\bk}|^2,\\
\nonumber
\fl&&\alpha_{2,\bk} = a_{\bk}^2+(A_{\bk}{+}J_{mfa}/2)^2 -(b_{\bk}{+}d_{\bk})^2-|B_{\bk}|^2,\\
\label{eq:beta}
\fl&&\beta_{1,\bk} = 4[a_{\bk}(A_{\bk}{-}J_{mfa}/2)-(b_{\bk}{-}d_{\bk})\Im B_{\bk}]^2
                 -(2\Re B_{\bk})^2[a_{\bk}^2-(b_{\bk}{-}d_{\bk})^2],\\
\nonumber
\fl&&\beta_{2,\bk} = 4[a_{\bk}(A_{\bk}{+}J_{mfa}/2)-(b_{\bk}{+}d_{\bk})\Im B_{\bk}]^2
                 -(2\Re B_{\bk})^2[a_{\bk}^2-(b_{\bk}{+}d_{\bk})^2],
\eea
within the notation of equations~(\ref{eq:notationAB},~\ref{eq:notationabd}), and the MFA-inspired
definition of $J_{mfa}= J_2+J_{\perp}-J_p$.

 The free terms in the first-order systems (see equation~(\ref{eq:coef_G1})) are determined by the
zero-order Green's functions, and thus the first-order quantities $G^{(1)}$ can be written down in terms
of the solution for the zero-order system \ref{eq:solutionG}, and the as-yet-unknown
quantities ${\rm v}_1$, ${\rm v}_2$, ${\rm v}_3$, and ${\rm v}_4$.

  To calculate ${\rm v}_{1,2,3,4}$ we use a relation that connects ${\rm v}$ and
the Green's functions $G^{(1)}(\bk,\tau=0^-)$, \emph{viz.}
\bea
\label{eq:V}
-{\rm v}_l&=&\frac 4N\sum_{\bk}G^{(1)}_{ll}(\bk,0^-),\qquad l=1,2,3,4.
\eea
 After the substitution of the solutions of the systems of equations
for the first-order Green's functions $G^{(1)}(\bk,\omega_m)$ in \ref{eq:solutionG1}
into the system of equations for ${\rm v}_{l}$  in equation~(\ref{eq:V}), the results are
found to be
\begin{eqnarray}
\label{eq:SSaz}
\fl&&{\rv}_1-{\rv}_2-{\rv}_3+{\rv}_4=\frac{\frac1{\beta}\sum\limits_m{\re}^{-{\ri}\omega_m0^-}{\cal N}^y(\omega_m)}
                  {2\eta - \frac1{\beta}\sum\limits_m{\re}^{-{\ri}\omega_m0^-}\left\{
                  {\ri}\omega_m{\cal D}^y(\omega_m)/\eta - 2{\cal Z}(J_{\perp}{+}J_2){\cal N}^y(\omega_m)\right\}},\\
\label{eq:SSbz}
\fl&&{\rv}_1+{\rv}_2-{\rv}_3-{\rv}_4=\frac{\frac1{\beta}\sum\limits_m{\re}^{-{\ri}\omega_m0^-}{\cal N}^z(\omega_m)}
                  {2\eta - \frac1{\beta}\sum\limits_m{\re}^{-{\ri}\omega_m0^-}\left\{
                  {\ri}\omega_m{\cal D}^z(\omega_m)/\eta - 2{\cal Z}(J_{\perp}{-}J_p){\cal N}^z(\omega_m)\right\}},
\end{eqnarray}
where
\bea
\nonumber
\fl&&{\cal N}^y(\omega_m)=\frac 4N\sum_{\bk}\bigg\{
              |G_{22}|^2{-}|G_{12}|^2{+}|G^-_{22}|^2{-}|G^-_{12}|^2
           {-}|G_{42}|^2{+}|G_{32}|^2{-}|G^-_{42}|^2{+}|G^-_{32}|^2\bigg\}~~,\\
\nonumber
\fl&&{\cal D}^y(\omega_m)=\frac 4N\sum_{\bk}\bigg\{
              |G_{22}|^2{-}|G_{12}|^2{-}|G^-_{22}|^2{+}|G^-_{12}|^2
           {-}|G_{42}|^2{+}|G_{32}|^2{+}|G^-_{42}|^2{-}|G^-_{32}|^2\bigg\}~~,\\
\fl&&\\
\nonumber
\fl&&{\cal N}^z(\omega_m)=\frac 4N\sum_{\bk}\bigg\{
              |G_{22}|^2{+}|G_{12}|^2{+}|G^-_{22}|^2{+}|G^-_{12}|^2
           {-}|G_{42}|^2{-}|G_{32}|^2{-}|G^-_{42}|^2{-}|G^-_{32}|^2\bigg\}~~,\\
\nonumber
\fl&&{\cal D}^z(\omega_m)=\frac 4N\sum_{\bk}\bigg\{
              |G_{22}|^2{+}|G_{12}|^2{-}|G^-_{22}|^2{-}|G^-_{12}|^2
           {-}|G_{42}|^2{-}|G_{32}|^2{+}|G^-_{42}|^2{+}|G^-_{32}|^2\bigg\}~~,\\
\fl&&
\eea
and all zero-order Green's functions $G(\bk,\omega_m)$ are given in \ref{eq:solutionG}.

 Now let us find the quantities which determine the linear response to a
magnetic field applied to the one of sublattices (\emph{e.g.}, see equation~(\ref{eq:iterpret1})).
 The longitudinal $z$ components of
the susceptibility in the characteristic representation are given by
\bea
\label{eq:chi_z}
\fl&&\chi^{\sigma^z\sigma^z}_{11}
=\frac{\partial \langle\sigma^z_1\rangle^f}{\partial f}\Big|_{f=0}={\rm v}_1,\qquad
\chi^{\sigma^z\sigma^z}_{12}
=\frac{\partial \langle\sigma^z_2\rangle^f}{\partial f}\Big|_{f=0}={\rm v}_2,\\
\nonumber
\fl&&
\chi^{\sigma^z\sigma^z}_{13}
=\frac{\partial \langle\sigma^z_3\rangle^f}{\partial f}\Big|_{f=0}={\rm v}_3,\qquad
\chi^{\sigma^z\sigma^z}_{14}
=\frac{\partial \langle\sigma^z_4\rangle^f}{\partial f}\Big|_{f=0}={\rm v}_4,
\eea
where the expansion of equation~(\ref{eq:explanation2}) was used.
The transverse $x$ and $y$ components of the susceptibility tensor are determined
in the terms of Green's functions to be given by
\bea
\label{eq:def}
\fl&& \chi^{\sigma^\alpha\sigma^{\alpha'}}_{11}\!\!=
 \frac 4N\sum_{i_1,i'_1}\int^{\beta}_0\!\!\langle T_{\tau}\sigma_{i_1}^{\alpha}(\tau)
 \sigma_{i'_1}^{\alpha'}(0)\rangle{\rm d}\tau,\quad
 \chi^{\sigma^{\alpha}\sigma^{\alpha'}}_{12}\!\!=
 \frac 4N\sum_{i_1,j_1}\int^{\beta}_0\!\!\langle T_{\tau}\sigma_{i_1}^{\alpha}(\tau)
 \sigma_{j_1}^{\alpha'}(0)\rangle{\rm d}\tau,\\
\nonumber
\fl&& \chi^{\sigma^\alpha\sigma^{\alpha'}}_{13}\!\!=
 \frac 4N\sum_{i_1,i'_2}\int^{\beta}_0\!\!\langle T_{\tau}\sigma_{i_1}^{\alpha}(\tau)
 \sigma_{i'_2}^{\alpha'}(0)\rangle{\rm d}\tau,\quad
 \chi^{\sigma^{\alpha}\sigma^{\alpha'}}_{14}\!\!=
 \frac 4N\sum_{i_1,j_2}\int^{\beta}_0\!\!\langle T_{\tau}\sigma_{i_1}^{\alpha}(\tau)
 \sigma_{j_2}^{\alpha'}(0)\rangle{\rm d}\tau,
\eea
where $\alpha=x,y$.
By substituting the solutions in \ref{eq:solutionG} into the
definition in equation~(\ref{eq:def}) for the transverse components of
susceptibility, we obtain the result given in \ref{trans_comp}.
 This result for the transverse components in the CR is {\emph{exactly the same}}
as the MFA calculations for the transverse components.

Then, using equations~(\ref{eq:CRtoINx})-(\ref{eq:CRtoINz}) the components of the susceptibility in the
initial coordinate system of equation~(1) below the transition temperature are found to be
\bea
\label{eq:SxSx}
\fl\chi^{x}&=&\frac 14 \frac 1{J_1 {+} J_2 {+} 2J_{\perp} {+} (J'_{\perp} {-} J_p)},\\
\label{eq:SySy}
\fl\chi^{y}&=&\frac 14 \frac {\sin^2(\theta)}
                                   {J_2 {-} J_3 {+} 2J_{\perp} {-} 2J_p}
           +\cos^2(\theta)[{\rm v}_1-{\rm v}_2-{\rm v}_3+{\rm v}_4],\\
\label{eq:SzSz}
\fl\chi^{z}&=&\frac 14 \frac {\cos^2(\theta)}
                                   {J_2 {+} J_3 {+} 2J_{\perp}}
           +\sin^2(\theta)[{\rm v}_1+{\rm v}_2-{\rm v}_3-{\rm v}_4]~~.
\eea

 These expressions for the components of susceptibility include the as-yet-unknown value
of the order parameter $\eta$.
 It can be found directly; from the definition of the Green's functions we have
\begin{eqnarray}
\label{eq:temp}
\fl G_{nn}(\tau=0^-)=\langle \sigma^-_n\sigma^+_n\rangle=\frac 12 - \eta,\quad
G_{nn}(\tau=0^-)=\frac 2N\sum_{\bk}G_{22}(\bk,\tau=0^-).
\end{eqnarray}
 Substituting $G_{22}(\bk,\omega)$ from equation~(\ref{eq:G22zero}), and performing the
summation on the Matsubara frequencies, the equation for the order parameter
turns out to be
\bea
\label{eq:sigma^z}
\fl\frac 1{\eta} =\frac12\frac 4N\sum_{\bk} \bigg\{\!\!&\hphantom{+}&
                                                \bigg(y_{1,\bk}{+}\frac{x_{1,\bk}}{\sqrt{\beta_{1,\bk}}}\bigg)
                                   \frac{2n(\varepsilon_{1,\bk}){+}1}{\omega_{1,\bk}}
                               +\bigg(y_{1,\bk}{-}\frac{x_{1,\bk}}{\sqrt{\beta_{1,\bk}}}\bigg)
                                   \frac{2n(\varepsilon_{2,\bk}){+}1}{\omega_{2,\bk}}\\
\nonumber\fl                              \!\!&+&\bigg(y_{2,\bk}{+}\frac{x_{2,\bk}}{\sqrt{\beta_{2,\bk}}}\bigg)
                                   \frac{2n(\varepsilon_{3,\bk}){+}1}{\omega_{3,\bk}}
                               +\bigg(y_{2,\bk}{-}\frac{x_{2,\bk}}{\sqrt{\beta_{2,\bk}}}\bigg)
                                   \frac{2n(\varepsilon_{4,\bk}){+}1}{\omega_{4,\bk}}\bigg\}~~,
\eea
where
\bea
\nonumber
\fl x_{1,\bk}&=&-2a_{\bk}[a_{\bk}(A_{\bk}{-}J_{mfa}/2)-(b_{\bk}{-}d_{\bk})\Im B_{\bk}],\qquad
       y_{1,\bk}=-(A_{\bk}{-}J_{mfa}/2),\\
\nonumber
\fl x_{2,\bk}&=&\hphantom{-}
                  2a_{\bk}[a_{\bk}(A_{\bk}{+}J_{mfa}/2)-(b_{\bk}{+}d_{\bk})\Im B_{\bk}],\qquad
       y_{2,\bk}=\hphantom{-}(A_{\bk}{+}J_{mfa}/2),\\
\nonumber
\fl n(\varepsilon_{l,\bk})&=&[\exp(\beta\varepsilon_{l,\bk})-1]^{-1},\qquad l=1,2,3,4.
\eea
 Since the order parameter (\emph {viz.}, the sublattice magnetization)
is temperature dependent, it follows that the spectrum of elementary
excitations (equation~(\ref{eq:varepsilon})) is also temperature dependent.

The N\'eel temperature at which $\eta$ vanishes within the adopted RPA
approximation is determined by
\bea
\label{eq:T_n}
\fl T_{\rm N} =\bigg[\frac1{2{\cal Z}}\frac 4N\sum_{\bk} \bigg\{\!\!&\hphantom{+}&
                                                \bigg(y_{1,\bk}{+}\frac{x_{1,\bk}}{\sqrt{\beta_{1,\bk}}}\bigg)
                                   \frac{1}{\omega^2_{1,\bk}}
                               +\bigg(y_{1,\bk}{-}\frac{x_{1,\bk}}{\sqrt{\beta_{1,\bk}}}\bigg)
                                   \frac{1}{\omega^2_{2,\bk}}\\
\nonumber\fl                              \!\!&+&\bigg(y_{2,\bk}{+}\frac{x_{2,\bk}}{\sqrt{\beta_{2,\bk}}}\bigg)
                                   \frac{1}{\omega^2_{3,\bk}}
                               +\bigg(y_{2,\bk}{-}\frac{x_{2,\bk}}{\sqrt{\beta_{2,\bk}}}\bigg)
                                   \frac{1}{\omega^2_{4,\bk}}\bigg\}\bigg]^{-1}.
\eea

 By putting $\eta\to 0$ we find that the $z$-component of the susceptibility
$\chi^{z}$ in equation~(\ref{eq:SzSz}) does not diverges at the N\'eel temperature,
whereas it diverges for the pure 2D model ($J_\perp=J_\perp^\prime=0$).
 At the N\'eel temperature all components of the susceptibility within the RPA
are equal to the MFA results, the latter of which are given in
equations~(\ref{eq:MFA_SxSx},~\ref{eq:MFA_SySy_Tn},~\ref{eq:MFA_SzSz_Tn}).

For completeness, we mention that the investigation of the model equation~(1) within
linear spin-wave (LSW) theory leads to the same structure of the susceptibility
expressions as we found within the RPA in equations~(\ref{eq:SxSx}-\ref{eq:SzSz}).
 The main difference between the results in RPA and LSW theory comes from the
calculation of the longitudinal components of the susceptibility in the CR.
 The spin-wave theory gives unity in the denominator of the expressions
in equations~(\ref{eq:SSaz},~\ref{eq:SSbz}), and $S=1/2$ instead of the order parameter $\eta$ everywhere
in the numerator ${\cal N}^y$ and ${\cal N}^z$.
 The transverse components of the susceptibility in the CR are equal
within the all of the MFA, RPA, and SW theories.

 When the temperature of the system is above the N\'eel temperature, $T_N$,
there still exists short-range antiferromagnetic order.
 To model such an order we follow reference \cite{Liu} and introduce a fictitious field $h$ pointing in
the direction of the sublattice magnetization, that is the $z$ direction
in the characteristic representation. To this end, the Hamiltonian
\begin{eqnarray}
\label{eq:H_para}
 H_h = H_{\rm CR} - h \sum_i\sigma^z_i - h \sum_j\sigma^z_j
\end{eqnarray}
is used, and the limit $h\to 0$ is taken after the calculation is
carried out.

  Above the N\'eel temperature, we define a (different) order parameter by
\begin{eqnarray}
\label{eq:y_def}
{\cal Y} = \lim_{h\to 0}(2{\cal Z}\eta/h).
\end{eqnarray}
 By a procedure similar to the above presented \cite{Tabunshchyk} (that is, the RPA scheme below $T_N$)
we have found the equation for the order parameter and all components of the magnetic
susceptibility in the paramagnetic phase.
It is then possible to show that paramagnetic version of the equation for the order
parameter in equation~(\ref{eq:sigma^z}) leads to
\bea
\label{eq:eq_y}
\fl \beta =\frac 1{2{\cal Z}}\frac 4N\sum_{\bk} \bigg\{\!\!&\hphantom{+}&
                                                \bigg(y_{1,\bk}{+}\frac{x_{1,\bk}}{\sqrt{\beta_{1,\bk}}}\bigg)
                                   \frac{1}{\omega^2_{1,\bk}}
                               +\bigg(y_{1,\bk}{-}\frac{x_{1,\bk}}{\sqrt{\beta_{1,\bk}}}\bigg)
                                   \frac{1}{\omega^2_{2,\bk}}\\
\nonumber\fl                              \!\!&+&\bigg(y_{2,\bk}{+}\frac{x_{2,\bk}}{\sqrt{\beta_{2,\bk}}}\bigg)
                                   \frac{1}{\omega^2_{3,\bk}}
                               +\bigg(y_{2,\bk}{-}\frac{x_{2,\bk}}{\sqrt{\beta_{2,\bk}}}\bigg)
                                   \frac{1}{\omega^2_{4,\bk}}\bigg\}.
\eea
where in \emph{all expressions} for $x_{1,2}$, $y_{1,2}$, $\beta_{1,2}$ and $\omega_{1-4}$ which determine equation~(\ref{eq:eq_y}) in the paramagnetic
phase, we use a new definition for $J_{mfa}$ (which we now call $\tilde J_{mfa}$) that reads
\bea
\label{eq:newJ_mfa}
\tilde J_{mfa}=J_2+J_{\perp}-J_p+\frac1{\cal Y}.
\eea
 The quantity ${\cal Y}$ approaches infinity as the temperature is lowered to
$T_{\rm N}$.
 Indeed, putting ${\cal Y}\to\infty$ in equation~(\ref{eq:eq_y}) we find the temperature
at which ${\cal Y}$  diverges, which is identically equal to the N\'eel temperature.

 We have found (see below for numerical results) that for model~equation~(1) all components of susceptibility are
continuous at the N\'eel point within the RPA.

\section{Numerical Results}

 In this section we present the result of numerical calculations
of the system modelled by the Hamiltonian of equation~(1) based on the above-presented
analytical formulae.

\subsection{Parameters regimes}

 Firstly, let us consider the set of model parameters that appears in the
Hamiltonian of equation~(1), \emph{viz.} in-plane parameters $J$, $d$
and $\Gamma$, and out-of-plane parameters $J_{\perp}$ and
$J'_{\perp}$.
 The in-plane parameter $d$ that describes the antisymmetric DM interaction and parameters
$\Gamma_{1,2,3}$ that give the pseudo-dipolar anisotropy, are of
order $10^{-2}$ and $10^{-4}$ respectively in units of $J$ \cite{Aharony,Koshibae},
and it has been shown that the only combination from
the pseudo-dipolar terms that affects the behaviour of the system is
$\Delta\Gamma\equiv\Gamma_1-\Gamma_3$ \cite{Tabunshchyk,Neto}.
 Thus, the in-plane part of the model, that is equations~(\ref{eq:H_DMa},\ref{eq:H_DMb}),
can be completely described by the AFM Heisenberg model with the DM antisymmetric
exchange interaction ${\bf D}$ and XY-like pseudo-dipolar anisotropy given by $\Delta\Gamma$.

 In order to examine the behaviour of the system with respect to the out-of-plane
parameters, we introduce the combination
\bea
\Delta J_{\perp}{\equiv}J_{\perp}-J'_{\perp}~~,
\eea
that describes the interplanar anisotropy interaction between nearest-neighbour spins
which we refer to as the net interplanar coupling, and the combination
\bea
\tilde{J}_{\perp}{\equiv}J_{\perp}{+}J'_{\perp}~.
\eea
 In our calculations we take $\Delta J_{\perp}$ to be of the order $10^{-5}-10^{-4}$ in units of $J$
(see \cite{Johnston} and references therein).

\begin{figure}[h]
\begin{center}
 \epsfxsize 0.47\textwidth\epsfbox{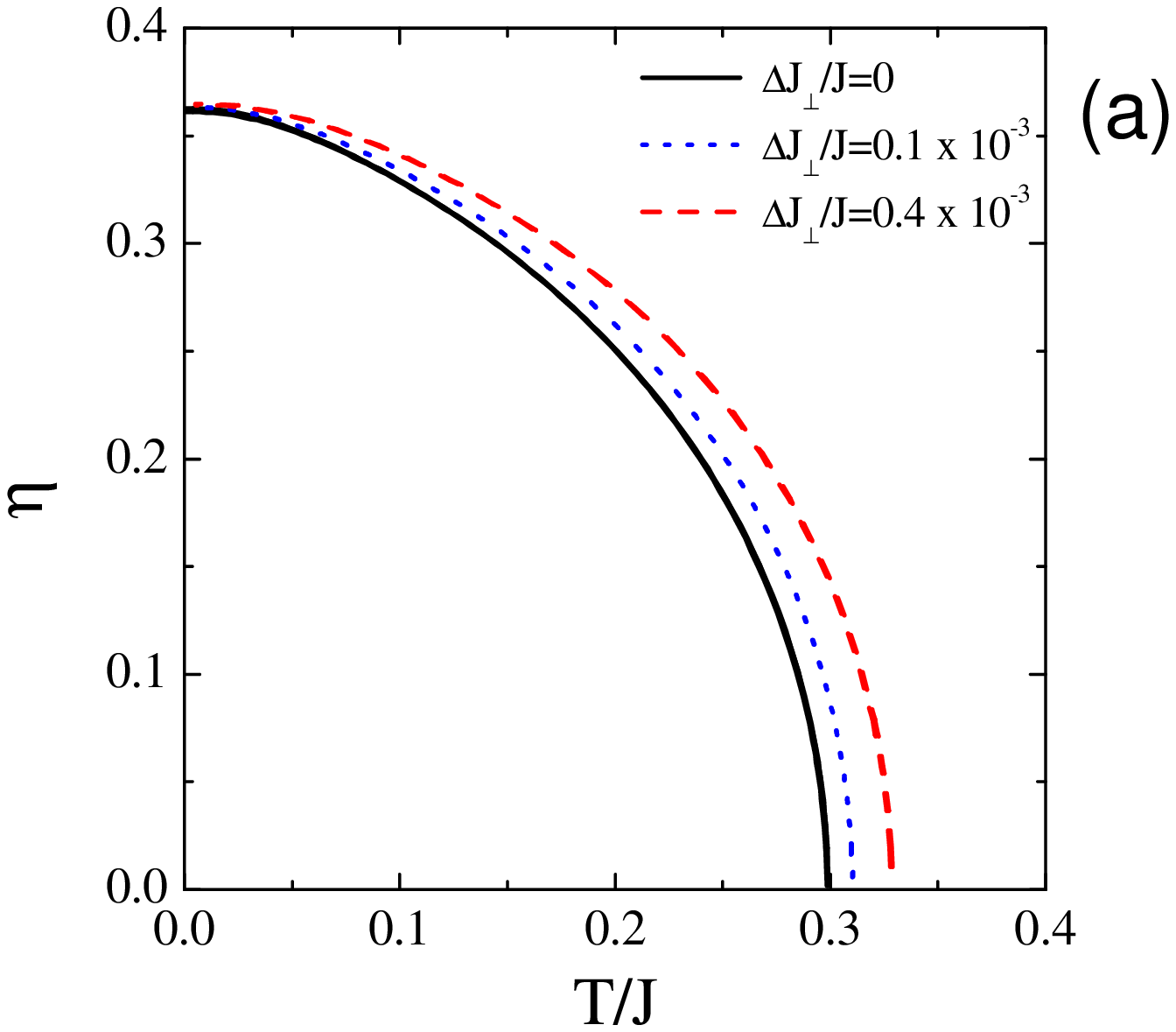}\quad
 \epsfxsize 0.47\textwidth\epsfbox{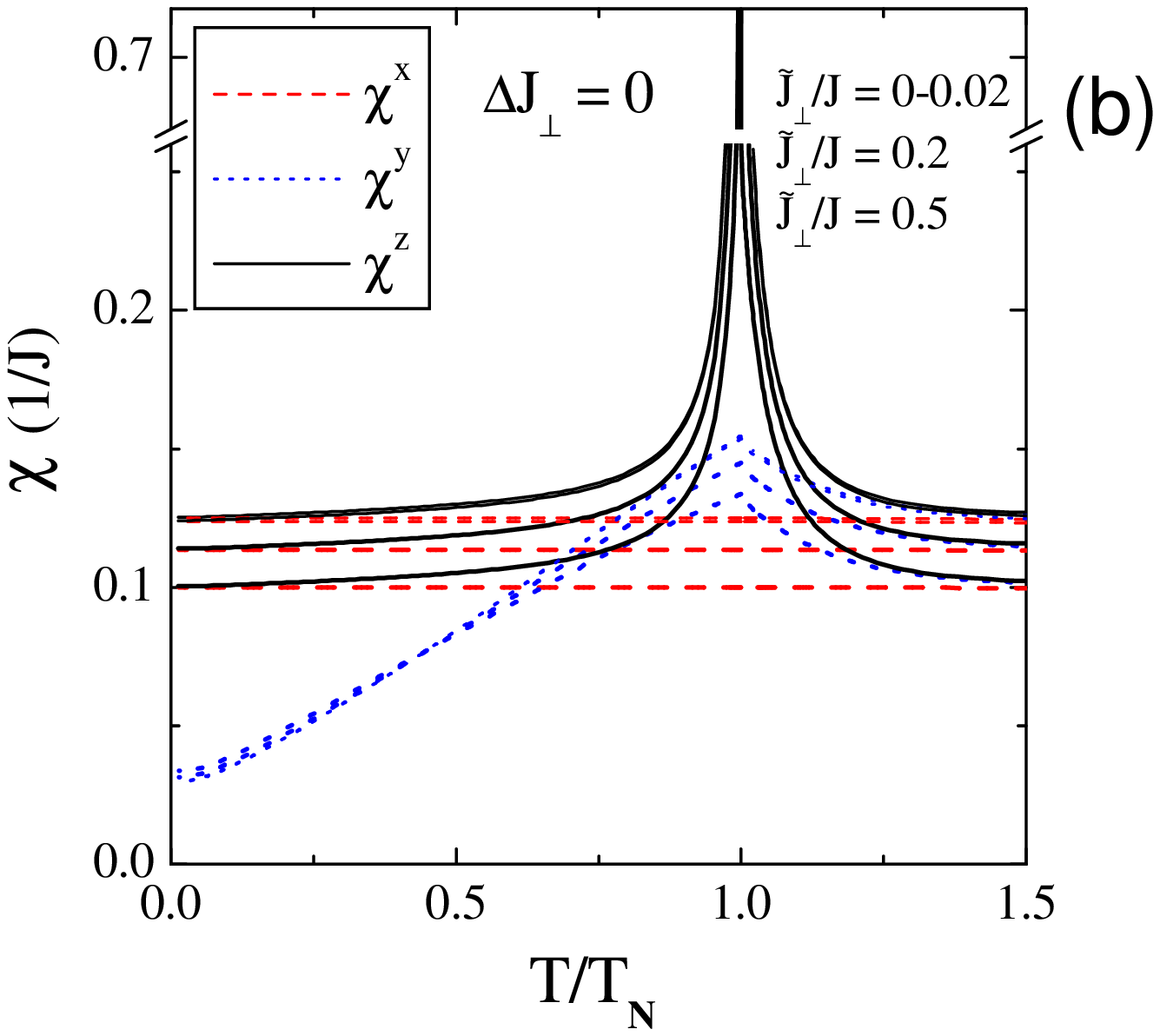}
\end{center}
\caption{\label{Fig01}(Colour online) The (a) order parameter \emph{vs.} $T/J$ for the dif\-f\-e\-r\-ent
                                      values of
                                      $\Delta J_{\perp}$: $\Delta J_{\perp}{=}0$ -- black solid line,
                                      $\Delta J_{\perp}{=}0.1{\times} 10^{-3}J$ -- blue dotted line,
                                      $\Delta J_{\perp}{=}0.4{\times} 10^{-3}J$ -- red dashed line,
                                     and each line consists of five data sets with $\tilde{J}_{\perp}$
                                     varying from zero up to $0.5J$.
                                      The (b) susceptibility, in units of $1/J$, {\emph vs.} $T/T_N$ for
                                      $\Delta J_{\perp}{=}0$, and for the following values of $\tilde{J}_{\perp}$:
                                      $\tilde{J}_{\perp}{=}0$ up to $0.02$ --- upper plot (many curves are superimposed on top of one another),
                                      $\tilde{J}_{\perp}{=}0.2$ --- middle plot,
                                  and $\tilde{J}_{\perp}{=}0.5$ --- lower plot.
                                      In both figures (a) and (b) we have fixed $d/J=0.02$ and
                                      $\Delta\Gamma/J=0.42\times10^{-3}$.}
\end{figure}

 In this subsection we focus on the behaviour of order parameter $\eta$, N\'eel temperature $T_{\rm N}$,
and susceptibility $\chi$ with respect to the parameter $\tilde{J}_{\perp}$ within the RPA method (we present a
detailed consideration of the dependence on $\Delta J_{\perp}$ in a subsequent subsection).
 Firstly, we find that the order parameter and the N\'eel temperature are almost independent of the
$\tilde{J}_{\perp}$ within a wide range of the model parameters.
 In figure~\ref{Fig01} we show two representative plots for the order parameter
and the susceptibility for certain  values of the in-plane parameters.
 \emph{In each line} of the figure~\ref{Fig01}a (that is solid, dotted, and dashed lines)
we have simultaneously plotted five data sets, each with different values of the parameter
$\tilde{J}_{\perp}$ that has been varied from zero up to $0.5J$
($\tilde{J}_{\perp}/J=0, 0.01, 0.1, 0.2, 0.5$).
 As one can see, for such a wide range of the parameter $\tilde{J}_{\perp}$ there is virtually no
difference of the absolute values of the N\'eel temperature and the order parameter, whereas
the relatively small changes of the net interplanar coupling $\Delta J_{\perp}$ in
figure~\ref{Fig01}a strongly affects these quantities.

In case of the susceptibility, its dependence on $\tilde{J}_{\perp}$ differs from that
discussed above for the order parameter $\eta$.
 Now, as is seen in figure~\ref{Fig01}b, the parameter $\tilde{J}_{\perp}$ generates the constant
shift in the $\chi^x$ and $\chi^z$ components of susceptibility as well as the constant shift
in the $\chi^y$ near the N\'eel temperature and within the paramagnetic region.
 However, for the reasonable values of the out-of-plane model parameters (that is $J_{\perp},\;
J'_{\perp} < 10^{-3}J$) the parameter $\tilde{J}_{\perp}$ does
\emph{not} affect on the behaviour of the susceptibility with temperature.

 This latter result for the susceptibility can be shown in various limits from the above
analytical results by taking into account the small magnitude of
$\Delta J_{\perp}$ and the in-plane parameters $d$ and $\Delta \Gamma$ with respect to $J$.
 In the zero temperature limit one can write down the susceptibility in the following form:
\bea \label{eq:Zero_SxSz} \fl\chi^{x,z}_{T\to 0}
                       \approx\frac 14 \frac 1{2J + \tilde{J}_{\perp}},\quad
\chi^{y}_{T\to 0}
                       \approx\frac 1{32} \frac 1{\{J - \tilde{J}_{\perp}/4\}^2}
                              \frac {d^2}{\Delta\Gamma+2\Delta J_{\perp}}.
\label{eq:Zero_SySy} \eea
Also, near the N\'eel temperature one finds
\bea \fl\chi^{y}_{T\to T_{\rm N}}
                     &\approx&\frac 1{32} \frac 1{\{J - \tilde{J}_{\perp}/4\}^2}
                               \frac {d^2}{\Delta\Gamma+2\Delta J_{\perp}}
                              +\frac 14\frac 1{2J +\tilde{J}_{\perp}},
\label{eq:Neel_SySy}\\
\fl\chi^{z}_{T\to T_{\rm N}}
                     &\approx&\frac 14 \frac 1{2J + \tilde{J}_{\perp}}+\frac 18 \frac 1{\tilde{J}_{\perp}}.
\label{eq:Neel_SzSz} \eea
 Almost everywhere within the above presented equations~(\ref{eq:Zero_SxSz}-\ref{eq:Neel_SzSz})
one can ignore the contribution of $\tilde{J}_{\perp} < 10^{-3}J$ with
respect to $J$;
only the $z$-component of the susceptibility $\chi^z$ at the N\'eel temperature is
strongly affected by $\tilde{J}_{\perp}$, as shown in equation~(\ref{eq:Neel_SzSz}).
 In fact, the upper plot in figure~\ref{Fig01}b consists of data sets with the
different values of the parameter $\tilde{J}_{\perp}$ over
the range of zero up to $0.02J$, but these differing plots can not be distinguished from
one another.

 Similarly, the parameter $\tilde{J}_{\perp}$ can be ignored in the expressions for the
spin-wave gaps, as can be clearly seen from the following
approximate formulae
\bea \label{eq:Omega1_appr}
\omega_{1,\bk\to
0}^2&\approx&\bigg\{J-\frac{\tilde{J}_{\perp}}2\bigg\}
\bigg(\Delta J_{\perp}+\frac d{\sqrt{2}}\theta-\big\{J-\frac{\tilde{J}_{\perp}}2\big\}\theta^2\bigg),\\
\label{eq:Omega2_appr} \omega_{2,\bk\to
0}^2&\approx&\bigg\{J+\frac{\tilde{J}_{\perp}}2\bigg\}
\bigg(\frac d{\sqrt{2}}\theta-\big\{J-\frac{\tilde{J}_{\perp}}2\big\}\theta^2\bigg),\\
\label{eq:Omega3_appr} \omega_{3,\bk\to
0}^2&\approx&\bigg\{J-\frac{\tilde{J}_{\perp}}2\bigg\}
(\Delta\Gamma/2 +\Delta J_{\perp}),\\
\label{eq:Omega4_appr} \omega_{4,\bk\to
0}^2&\approx&\bigg\{J+\frac{\tilde{J}_{\perp}}2\bigg\}
\Delta\Gamma/2.
\eea
 Therefore, one can conclude that the affect of the parameter $\tilde{J}_{\perp}$
on the physics of the model is negligibly small.
 Consequently, without further trepidations the system can be studied using a fixed and
representative value of this parameter, \emph{e.g.} $\tilde{J}_{\perp}=10^{-3}J$, without having to
be concerned with it changing our results.

 At the end of this subsection we discuss briefly the case of isotropic interplanar coupling
($\Delta J_{\perp}=0$).
 In such a case the only difference in the susceptibility, with respect to a pure 2D model \cite{Tabunshchyk},
is the finite value of $\chi^z$ at the N\'eel temperature.
 Then, only the in-plane anisotropies are responsible for the anisotropic magnetic properties of such a system,
\emph{viz.} the behaviour of susceptibility, order parameter, and
spin-wave excitations.
 We emphasize the perhaps expected result, that for the 3D case with isotropic interplanar coupling,
due to the frustration of interplanar coupling within the body-centred lattice, the
effects of 2D quantum fluctuations and short-range correlations are very important, whereas the interplanar
coupling is not.

\subsection{N\'eel temperature and spin-wave excitations}

 Now we present the results of our numerical investigations of the
N\'eel temperature and the spin wave excitations and their dependence on
the parameters of the in-plane anisotropies $d/J$, $\Delta\Gamma/J$, and the
out-of plane anisotropy $\Delta J_{\perp}/J$.

\begin{figure}[h]
\begin{center}
 \epsfxsize 0.47\textwidth\epsfbox{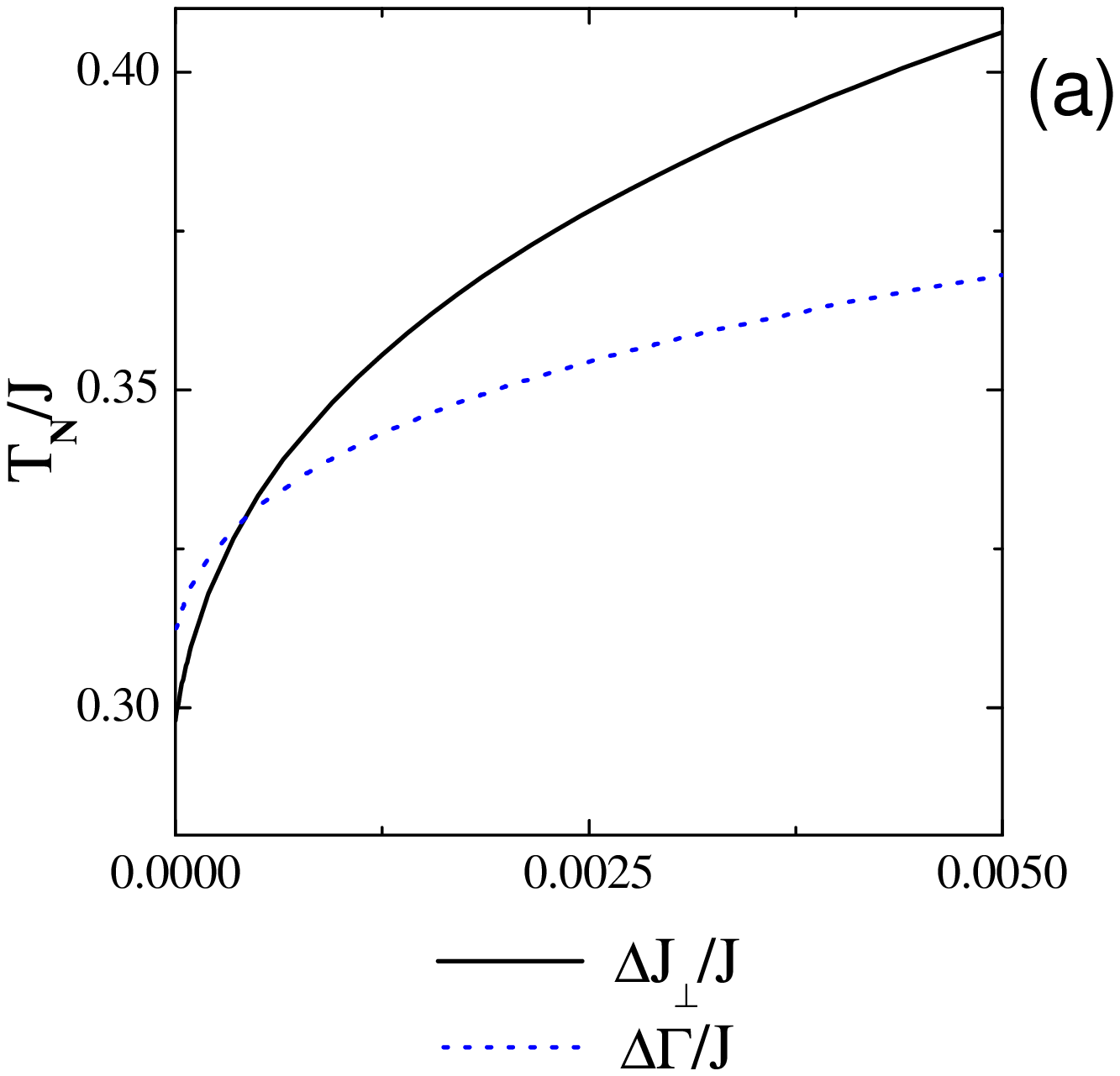}\quad
 \epsfxsize 0.47\textwidth\epsfbox{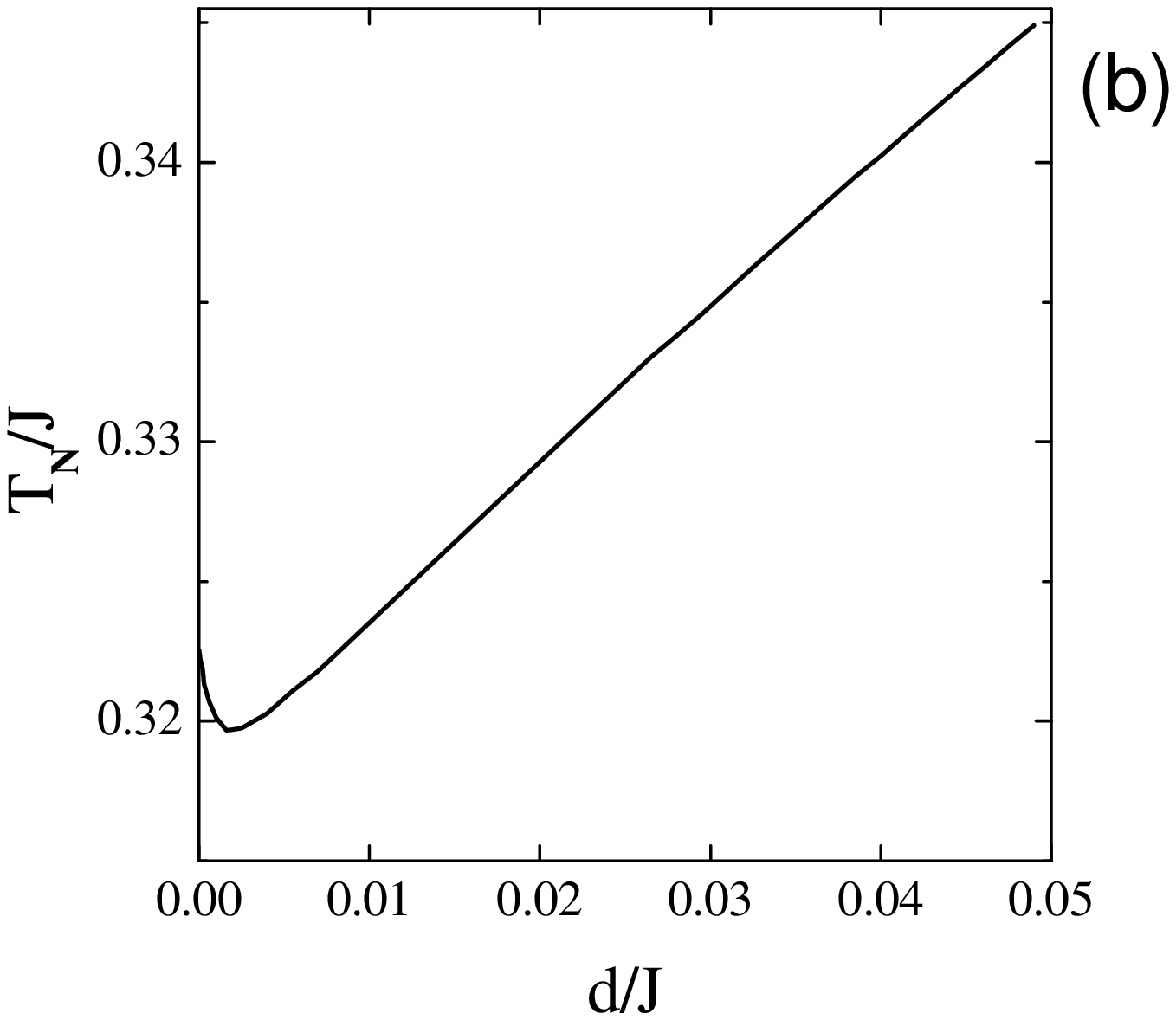}
\end{center}
\caption{\label{Fig02}(Colour online) The N\'eel temperature
$T_{N}$, in units of $J$, \emph{vs.} (a) $\Delta J_{\perp}/J$ --
black solid line (fixed $\Delta \Gamma /J=0.42\times10^{-3}$) and
$\Delta\Gamma/J$ -- blue dotted line (fixed $\Delta J_{\perp}/J=0.42\times10^{-3}$), both for fixed
$d/J=0.02$, as well as \emph{vs.} (b) the  DM parameter $d/J$ (fixed
$\Delta\Gamma/J=0.42\times10^{-3}$ and $\Delta
J_{\perp}/J=0.42\times10^{-3}$).}
\end{figure}

 Figure~\ref{Fig02}a shows the N\'eel temperature obtained within the RPA scheme as a function
of both $\Delta\Gamma/J$ and $\Delta J_{\perp}/J$.
 We found that the transition temperature $T_N$ depends on both the in-plane XY-like pseudo-dipolar
anisotropy parameter $\Delta\Gamma$ and the out-of-plane anisotropy parameter $\Delta J_{\perp}$,
and changes of the same order ($\sim 10^{-4}J$) in $\Delta J_{\perp}$ and/or in $\Delta\Gamma$
produces considerable changes in the N\'eel temperature, $T_N$. Further, the dependence of
the N\'eel temperature on the DM parameter is shown in figure~\ref{Fig02}b: $T_N$ decreases
as $d$ increases for small $d$, but for larger values of the DM interaction, i.e. $d\gg\Delta
J_{\perp},\Delta\Gamma$, the N\'eel temperature $T_N$ increases nearly linearly with $d$.
 The transition temperature into the long-range ordered state, $T_N$, increases as the
parameter $\Delta J_{\perp}$ increases since the net interplanar coupling that
each spin feels favours to the AFM state.

\begin{figure}[h]
\begin{center}
 \epsfxsize 0.47\textwidth\epsfbox{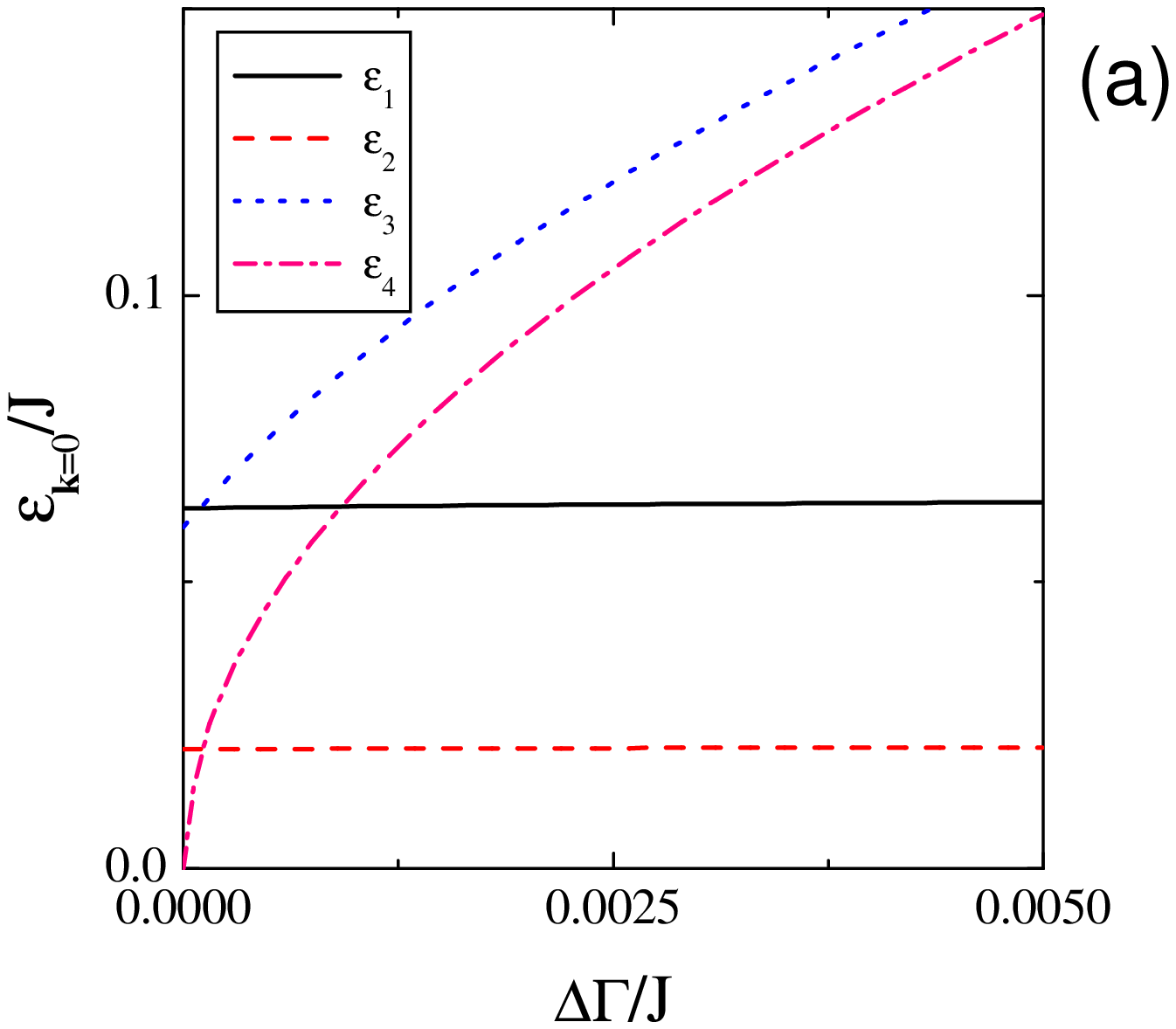}\quad\quad
 \epsfxsize 0.47\textwidth\epsfbox{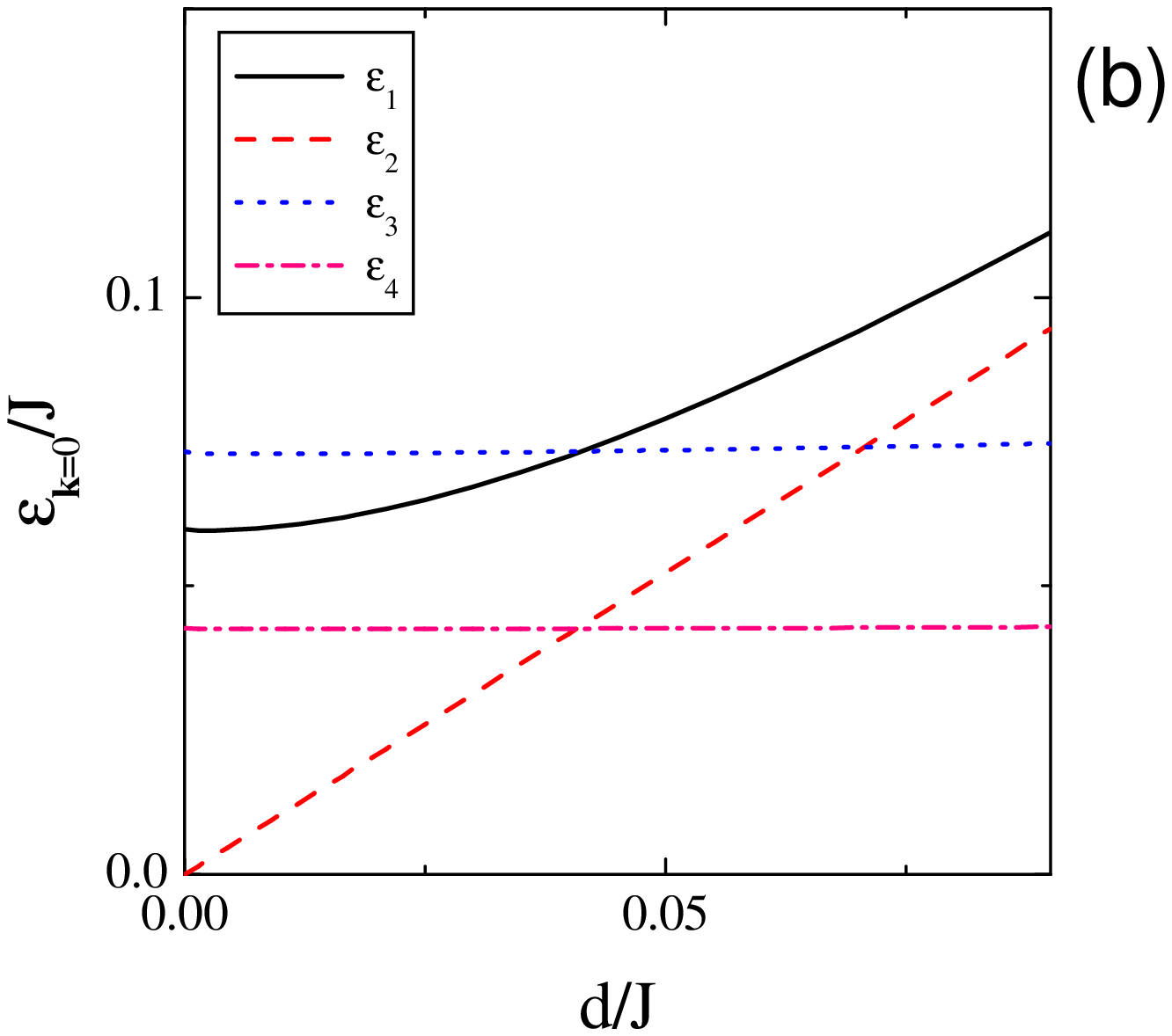}\\
 \epsfxsize 0.47\textwidth\epsfbox{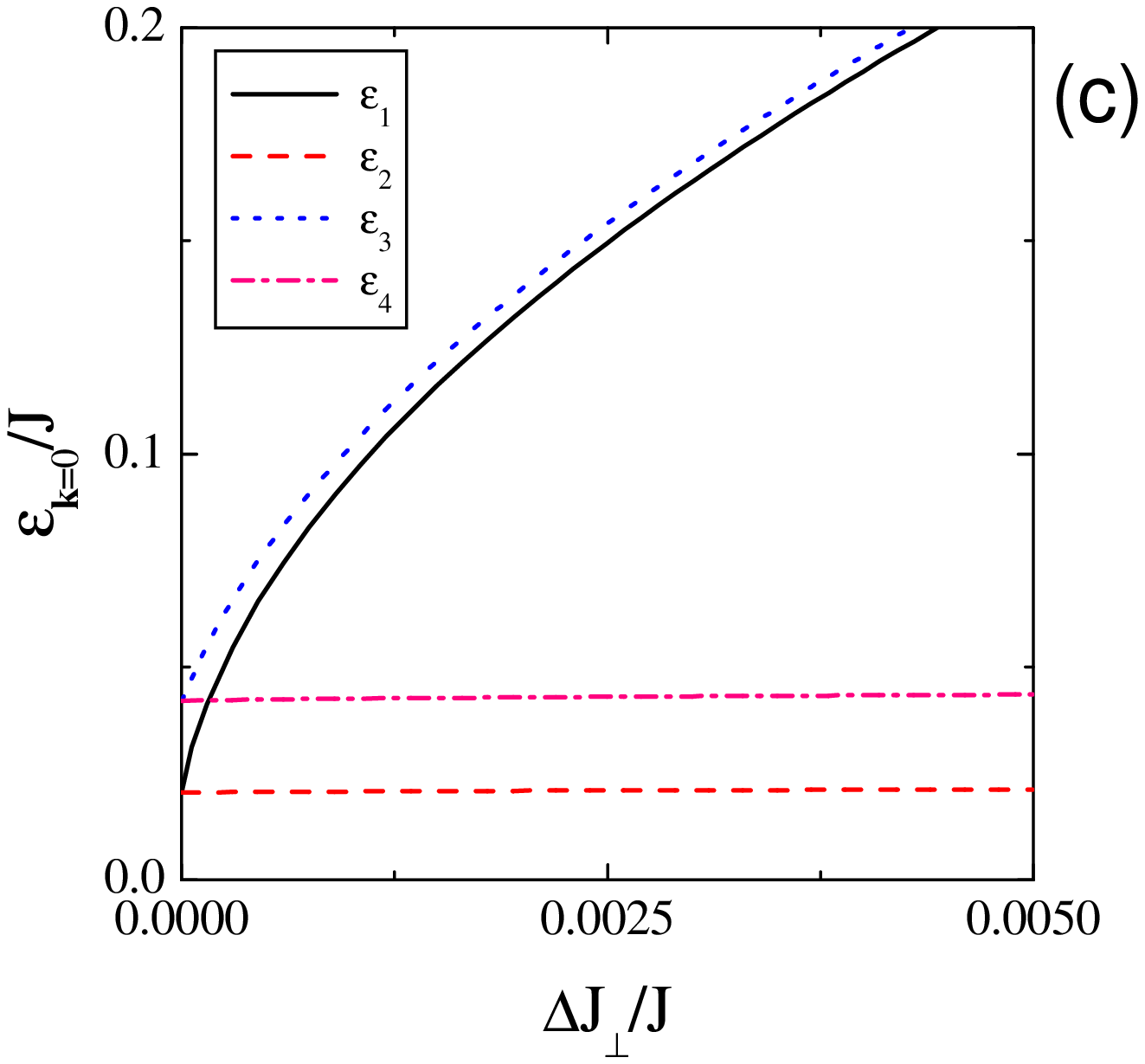}
\end{center}
\caption{\label{Fig03}(Colour online) The energy gaps, in units of $J$, as function of\\
(a) XY-like anisotropy parameter $\Delta\Gamma/J$
         ($d/J=0.02$, $\Delta J_{\perp}/J=0.42\times10^{-3}$),\\
     (b) DM parameter $d/J$
         ($\Delta\Gamma/J=0.42\times10^{-3}$, $\Delta J_{\perp}/J=0.42\times10^{-3}$),
and\\(c) out-of-plane anisotropy parameter $\Delta J_{\perp}/J$
($d/J=0.02$, $\Delta\Gamma/J=0.42\times10^{-3}$). }
\end{figure}

 Figure~\ref{Fig03} shows the zero-temperature energy gaps in the long
wavelength limit as a function of in- and out- of plane anisotropy
parameters,
 and the resulting behaviours can be understood immediately from
equations~(\ref{eq:Omega1_appr})-(\ref{eq:Omega4_appr}).
 Two modes, $\varepsilon_1$ and $\varepsilon_2$, are almost
independent of $\Delta\Gamma$ (see figure~\ref{Fig03}a, and
equations~(\ref{eq:Omega1_appr},\ref{eq:Omega2_appr})), but they show
a strong dependency on the DM parameter, $d$, as seen in
figure~\ref{Fig03}b.
 Since the canted angle goes as
$\theta\approx(d/\sqrt{2})/(2J)$, the modes $\varepsilon_1$,
$\varepsilon_2$ are nearly linear in $d$.
 In the limit of zero DM interaction the mode $\varepsilon_2$ goes to
zero and a Goldstone mode appears in the spin-wave spectrum,
while the mode $\varepsilon_1$ goes to a finite value, which is about
$2{\cal Z}\eta\sqrt{J\Delta J_{\perp}}$
(see equations~(\ref{eq:Omega1_appr},\ref{eq:Omega2_appr})).
 Two other modes in the spectrum, $\varepsilon_3$,
$\varepsilon_4$, are almost independent of the DM parameter of
anisotropy, while they vary strongly with $\Delta\Gamma$.
 In the limit $\Delta\Gamma=0$ the mode $\varepsilon_4$ goes to zero
and another Goldstone mode appears in the spectrum,
while the mode $\varepsilon_3$ goes to the finite value of about
$2{\cal Z}\eta\sqrt{J\Delta\Gamma/2}$
(equation~(\ref{eq:Omega3_appr},\ref{eq:Omega4_appr})).
 Since in the case of 3D model thermal fluctuations do not
destroy the long-range ordering for $T\ne 0$, the N\'eel
temperature does not go to zero when a Goldstone mode
$\varepsilon_2$ or $\varepsilon_4$ appears in the spectrum.

 Lastly, we note that the plots of figure~\ref{Fig03}c show that two modes
($\varepsilon_2$ and $\varepsilon_4$) are independent
of the net interplanar coupling $\Delta J_{\perp}$, and two modes
($\varepsilon_1$ and $\varepsilon_3$) demonstrate a square-root
dependency on $\Delta J_{\perp}$.
 When the net interplanar coupling goes to zero, $\Delta J_{\perp}=0$,
the two modes $\varepsilon_1$ and $\varepsilon_2$ become equal and
describe the in-plane mode in the spin-wave excitations \cite{Keimer}. Similarly,
the mode $\varepsilon_3$ coincides with $\varepsilon_4$ at $\Delta
J_{\perp}=0$ and they correspond to the out-of-plane magnon mode \cite{Keimer}.

\subsection{Susceptibility}

\begin{figure}[h]
\begin{center}
 \epsfxsize 0.47\textwidth\epsfbox{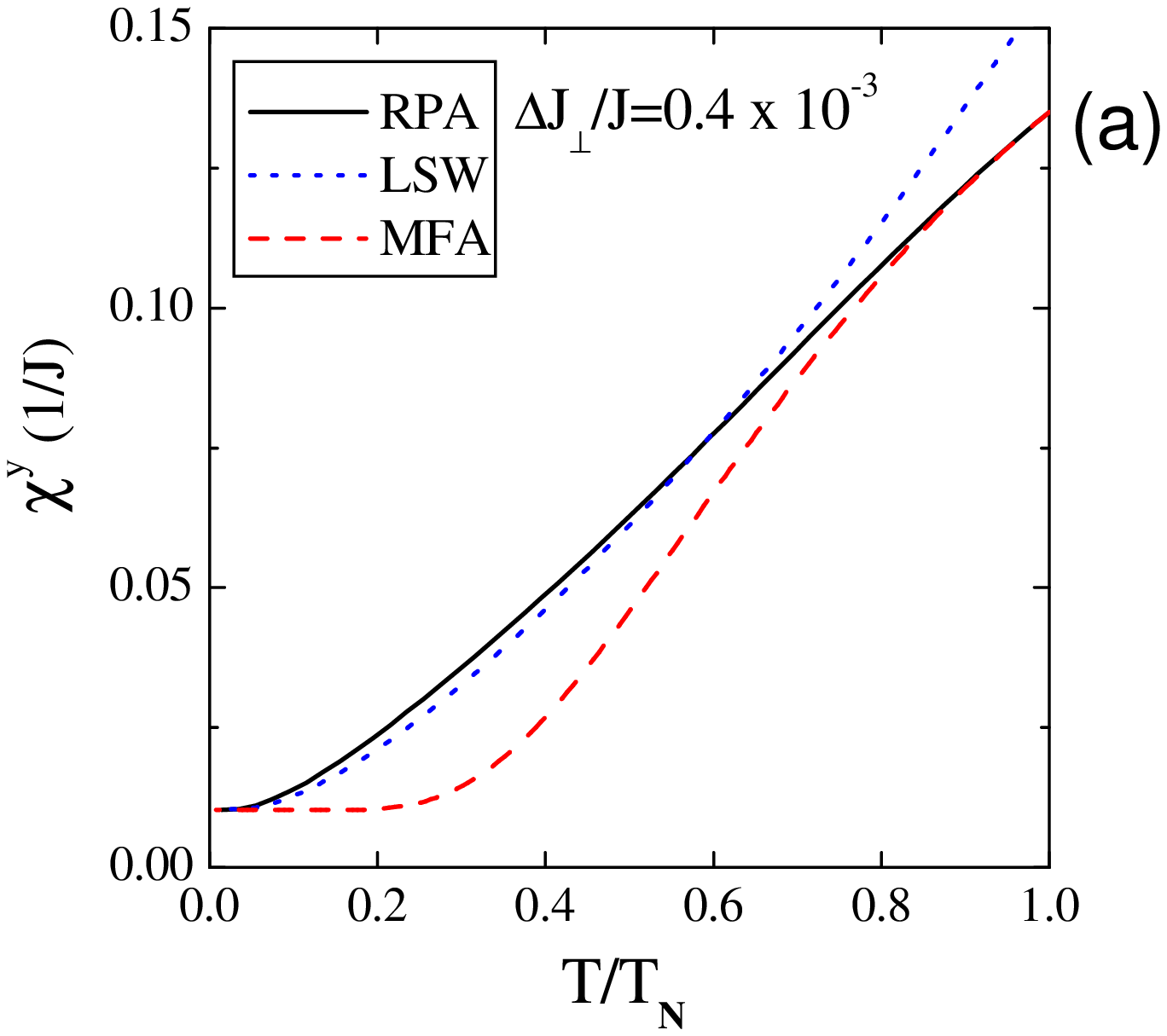}\quad
 \epsfxsize 0.47\textwidth\epsfbox{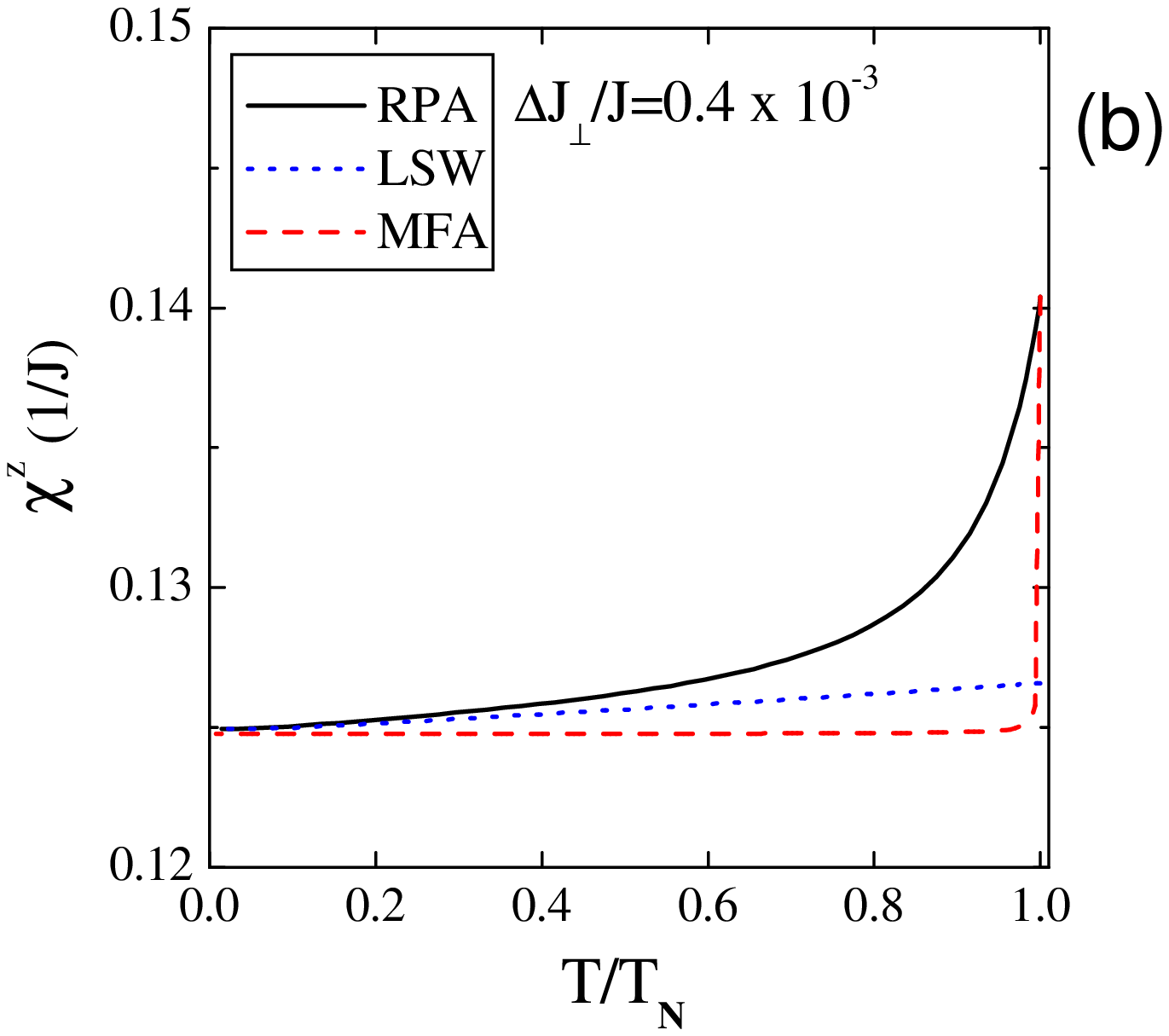}
\end{center}
\caption{\label{Fig05}(Colour online)
 The (a) $\chi^y$ and (b) $\chi^z$ components of the susceptibility, in
units of $1/J$ -- a
comparison of the RPA (black solid line), LSW (blue dotted line) and MFA (red dashed line)
results below $T_N$ (for $d/J=0.02$, $\Delta\Gamma/J=0.42\times10^{-3}$,
$\Delta J_{\perp}/J=0.4\times10^{-3}$).}
\end{figure}

 Now we consider the temperature behaviour of the susceptibility and
examine its dependency on
different values of the in-plane and out-of plane anisotropy parameters.
 Our results for the $y$, and $z$ components of the susceptibility within the
different approximation schemes, \emph{viz.} RPA, LSW theory, and MFA, are presented
in figure~\ref{Fig05} ($T<T_N$).
 We do not present the similar comparing for the $x$ component of susceptibility
because of the pure transverse component $\chi^x$ (see equation~(\ref{eq:SxSx}))
has the same value within all mentioned approximations (below the transition temperature).
 On the other hand, the longitudinal (in the characteristic  representation equations~
(\ref{eq:CRtoINy},\ref{eq:CRtoINz})) components of the susceptibility are involved in the equations
for the components $\chi^y$ and $\chi^z$ that leads to their different temperature behaviours within
the different approximation methods (see below).

 Our results for the $y$ component of the susceptibility, $\chi^y$, are shown in figure~\ref{Fig05}a.
 We find that at low temperatures the RPA analytical scheme, as was also found in pure 2D case \cite{Tabunshchyk},
is in good agreement with the linear spin-wave theory.
 Plots in figure~\ref{Fig05}a also show that RPA results agree with the MFA as one nears the transition
temperature $T_N$, and both RPA and MFA lead to the same magnitude of susceptibility at the N\'eel temperature.
 (It is worth noting here that transition temperature $T_N$ within the MFA approach, where
$T_N=J_2+J_{\perp}-J_p\approx J$, is almost independent of the anisotropy, in contrast to the
RPA scheme where $T_N$ is very sensitive to the anisotropy parameters (see figure~\ref{Fig02}).)
 One can also see that in the zero temperature limit all approximations used in the paper
go to the same value of susceptibility approximately given by equation~(\ref{eq:Zero_SxSz}).

 Figure~\ref{Fig05}b shows the $z$ component of the susceptibility $\chi^z$,
and again we obtain the good agreement between the RPA and LSW methods at low $T$,
and coincidence of all results in the zero temperature limit.
 On the other hand, in the vicinity of the transition temperature, the RPA scheme gives
qualitatively different behaviour of $\chi^z$ with respect to the MFA and LSW formalisms.
 Thus, we can answer one of the motivating questions of this study: Does the extension
of the model of reference \cite{Tabunshchyk} from 2D to 3D lead to a reduction of the strong effects of
quantum fluctuations? Indeed, the answer is no, and there are strong effects of
quantum fluctuations in our 3D Heisenberg model with the anisotropies.
 This statement is correct for a magnitude of the net interplanar coupling $\Delta J_{\perp}$ up to
$\sim 10^{-3}J$.

 Now let us find the correlation between the ratio of spin-waves modes of the magnon excitation
spectrum in the long wavelength limit, and the behaviour of the
components of susceptibility in the zero temperature limit
(similar to the correlation that we have found in the pure 2D
model \cite{Tabunshchyk}).
 Firstly, from the analytical results, equation~(\ref{eq:Zero_SxSz}), we obtain that the ratio between
the components of the susceptibility is given approximately by
\bea
\label{eq:ratio_chi}
\frac{\chi^{y}}{\chi^{x,z}}\bigg|_{T\to 0}\approx\frac{d^2}{4J(\Delta\Gamma+2\Delta J_{\perp})}.
\eea
 Next, by taking into account that the canted angle is $\theta\approx(d/\sqrt{2})/(2J)$ and
$\tilde{J}_{\perp}\ll J$, we can rewrite the expressions (\ref{eq:Omega1_appr})-(\ref{eq:Omega4_appr})
that specify the spin-wave gaps, which we write in a scaled form using $\varepsilon_l=2{\cal Z}\eta\omega_l$
as
\bea
\label{eq:Omega_appr_all}
\fl\omega_1^2 \approx J\Delta J_{\perp}+ d^2/8,\quad
   \omega_2^2 \approx d^2/8,\quad
   \omega_3^2 \approx J(\Delta\Gamma/2{+}\Delta J_{\perp}),\quad
   \omega_4^2 \approx J \Delta\Gamma/2.
\eea
 Thus, the ratio between the components of the susceptibility turns out to be
\bea
\label{eq:ratio_chi_Omega}
\frac{\chi^{y}}{\chi^{x,z}}\bigg|_{T\to 0}\approx\bigg(\frac{\varepsilon_2}{\varepsilon_3}\bigg)^2_{\bk\to 0}.
\eea
\begin{figure}[h]
\begin{center}
 \epsfxsize 0.47\textwidth\epsfbox{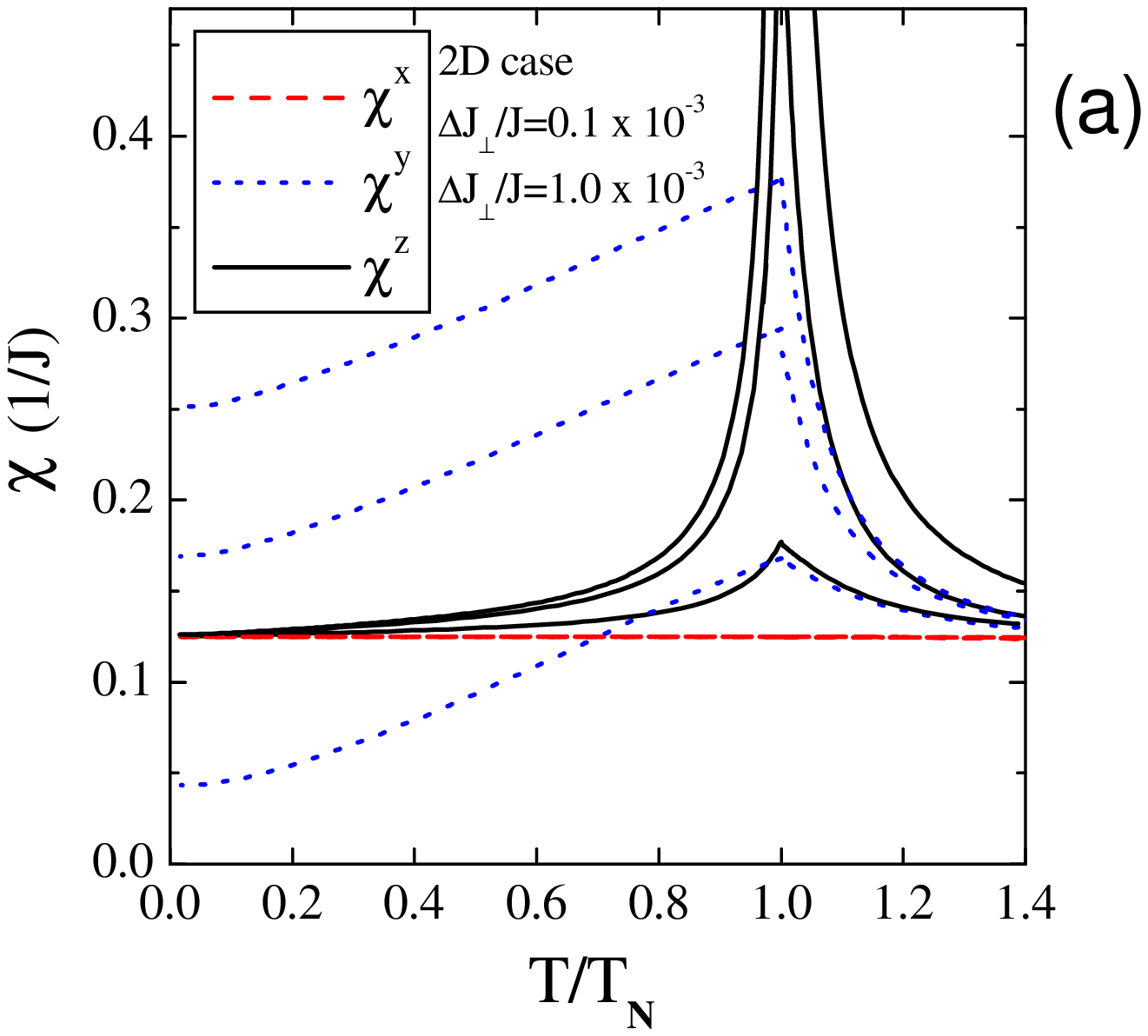}
 \epsfxsize 0.47\textwidth\epsfbox{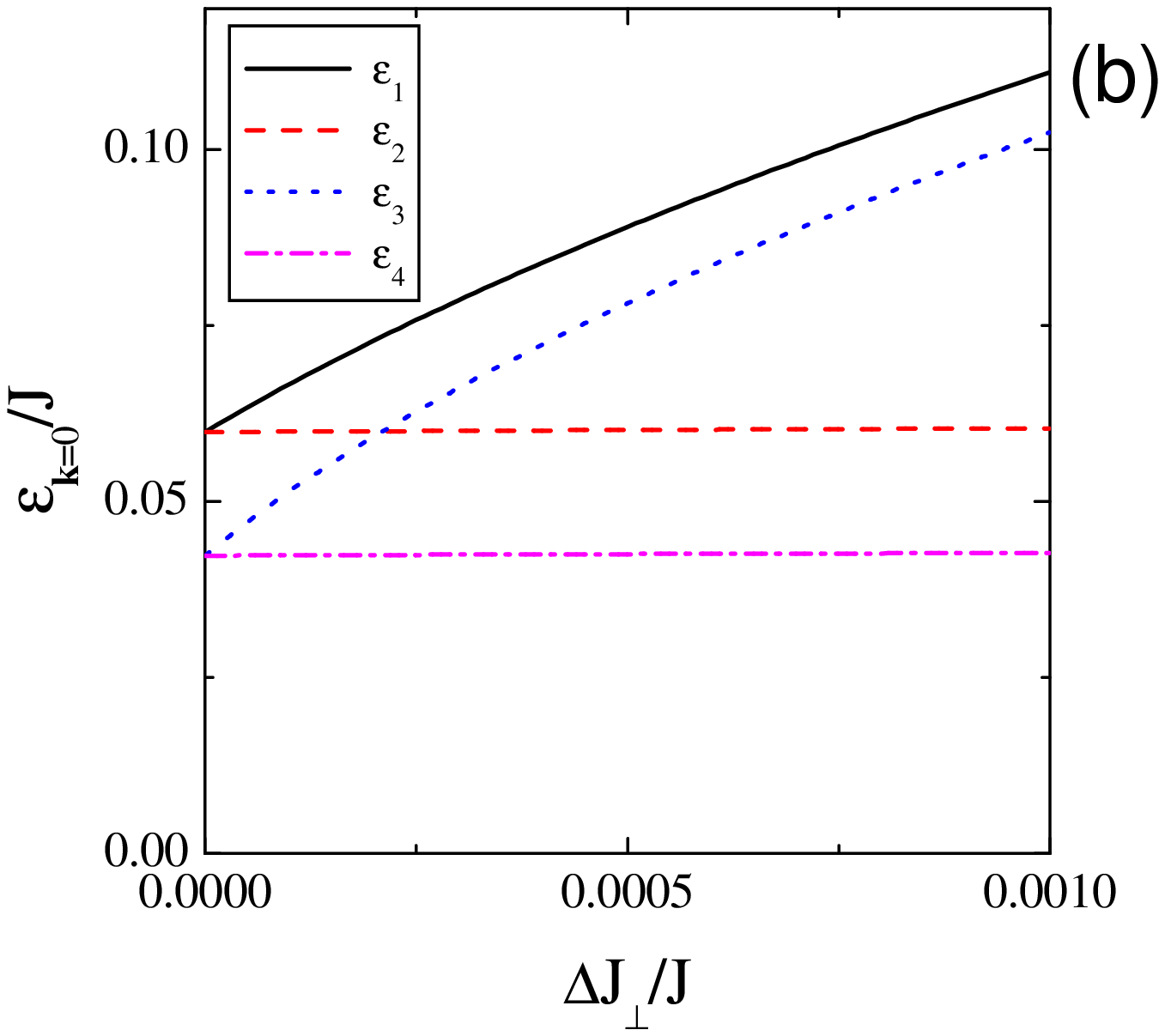}
\end{center}
\caption{\label{Fig06}(Colour online) (a) All three components of the susceptibility within the
RPA analytical scheme for different values of the interplanar anisotropy $\Delta J_{\perp}$: 2D result -- upper curve, $\Delta J_{\perp}/J=0.1\times 10^{-3}$ --
middle curve, and $\Delta J_{\perp}/J=1.0\times 10^{-3}$ -- top curve;
and the (b) $T=0$ energy gaps \emph{vs.} the interplanar anisotropy $\Delta J_{\perp}$, for
$d/J=0.058$, $\Delta\Gamma/J=0.42\times10^{-3}$.}
\end{figure}

  We also find that gap $\varepsilon_1$ is always greater than $\varepsilon_2$, and
$\varepsilon_3$ is always greater than $\varepsilon_4$, when $\Delta J_{\perp}=J_{\perp}-J'_{\perp}>0$
(indeed, that is the only case considered in this paper - see the discussion of figure 1 in section 2).
 As was mentioned above, the two modes $\varepsilon_1$, $\varepsilon_2$ describe the in-plane modes
of the spectrum, while the modes $\varepsilon_3$, $\varepsilon_4$ describe the out-of-plane
spin-wave excitations.
 Therefore, we find that the observed ratio between the $x$ and $y$ components $\chi^x<\chi^y$
(in the $T=0$ limit) \cite{Lavrov}, in any of the MFA, LSW theory, or the RPA,
takes place only if the spin-wave gaps have the following hierarchy:
\bea
\label{eq:hierarchy}
\varepsilon_1>\varepsilon_2>\varepsilon_3>\varepsilon_4,
\eea
\emph{i.e.} the in-plane modes ($\varepsilon_{1,2}$) are greater than the out-of-plane
ones ($\varepsilon_{3,4}$).

 The described situation is presented in figure~\ref{Fig06} -- they show
the susceptibility for the different values of the interplanar parameter $\Delta J_{\perp}$.
 The upper curve was obtained for the pure 2D case and corresponds to the situation with
the observed order of the susceptibility components $\chi^x<\chi^y$ and the following
ordering of the gaps $\varepsilon_1=\varepsilon_2 > \varepsilon_3=\varepsilon_4$.
 By increasing the magnitude of the interplanar parameter $\Delta J_{\perp}$
two modes $\varepsilon_1$ and $\varepsilon_3$ increase and the hierarchy of the gaps
(\ref{eq:hierarchy}) remains unchanged (figure~\ref{Fig06}b).
 As we can see from the middle curve in figure~\ref{Fig06}a, the ratio
$\chi^y/\chi^x$ decreases as the ratio between the gaps $\varepsilon_2/\varepsilon_3$
decreases.
 The magnitude of the out-of-plane mode $\varepsilon_3$ becomes equal to the in-plane one $\varepsilon_2$
when $\Delta J_{\perp}/J\approx \frac 12 \big(\frac 14 (d/J)^2-\Delta \Gamma/J \big)
\approx 2.1\times 10^{-4}$, where the ratio $\chi^y/\chi^x$ goes to unity. A further increasing
of $\Delta J_{\perp}$ changes the ratio between the modes $\varepsilon_2/\varepsilon_3$, and according
to equation~(\ref{eq:ratio_chi}) changes the order of the susceptibility components $x,z$ and $y$
at zero temperature
(see the lowest curve in figure~\ref{Fig06}a and the corresponding value of the gaps at $\Delta J_{\perp}=0.001$
in figure~\ref{Fig06}b).

\begin{figure}[h]
\begin{center}
 \epsfxsize 0.47\textwidth\epsfbox{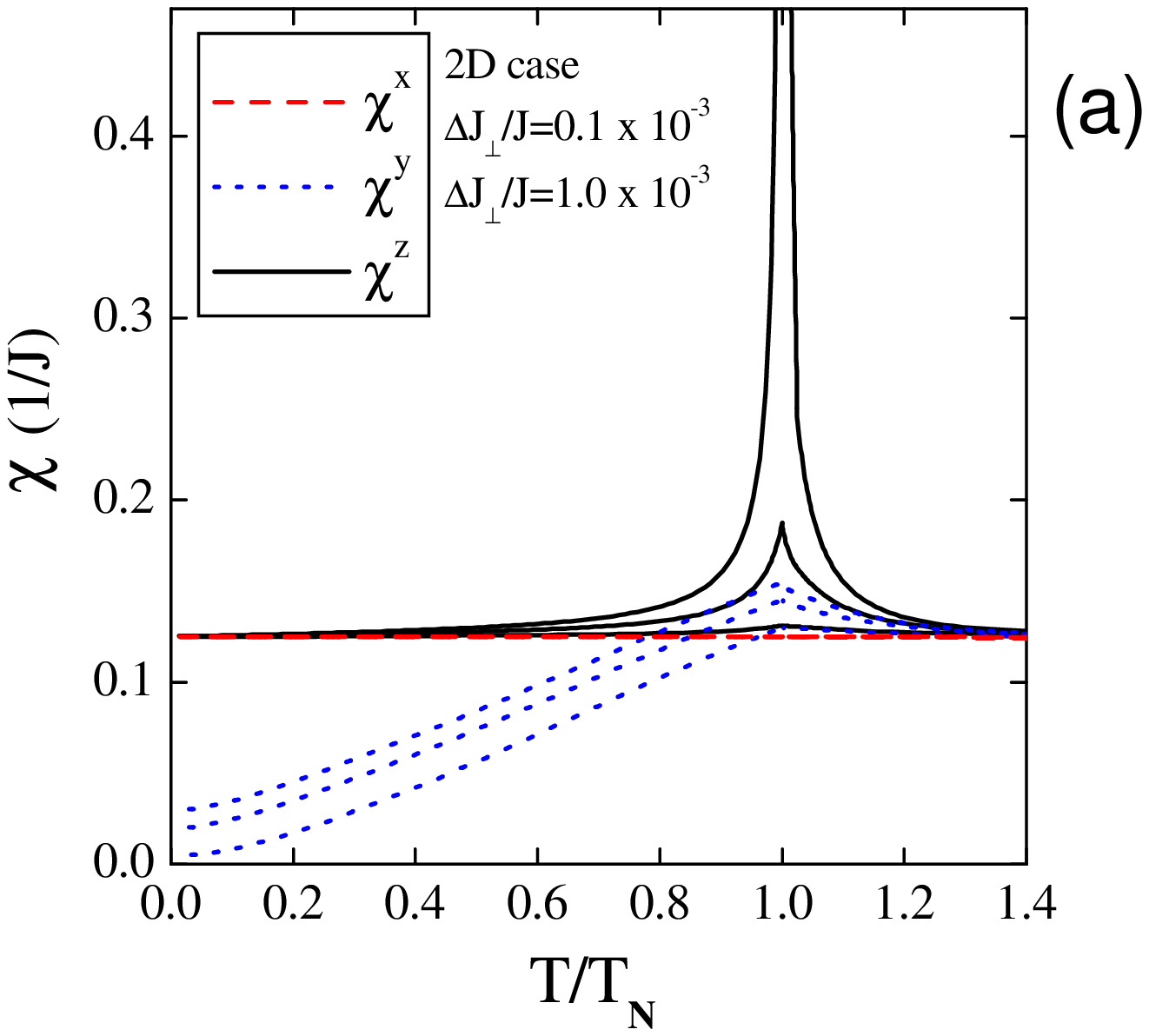}
 \epsfxsize 0.47\textwidth\epsfbox{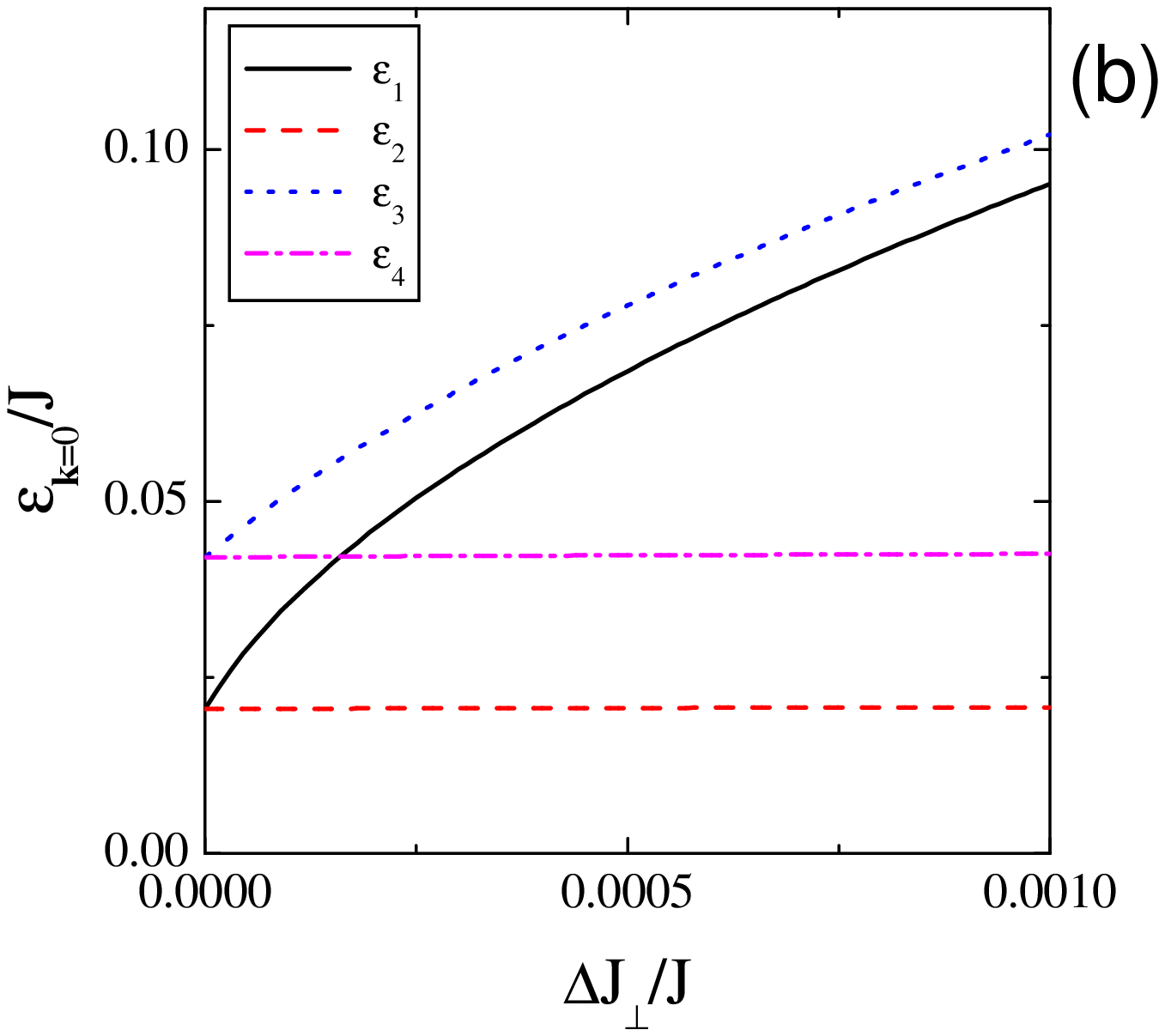}
\end{center}
\caption{\label{Fig07}(Colour online) (a) All three components of the susceptibility within the
RPA analytical scheme for
different values of the interplanar anisotropy $\Delta J_{\perp}$: 2D result -- upper curve, $\Delta J_{\perp}/J=0.1\times 10^{-3}$ --middle curve, and $\Delta J_{\perp}/J=1.0\times 10^{-3}$ -- top curve;
and the (b) $T=0$ energy gaps \emph{vs.} the interplanar anisotropy $\Delta J_{\perp}$, for
$d/J=0.02$, $\Delta\Gamma/J=0.42\times10^{-3}$.}
\end{figure}

 For completeness, in figure~\ref{Fig07} we present the susceptibility and behaviour of the gaps {\emph{vs.}}
$\Delta J_{\perp}$ for smaller magnitudes of the in-plane anisotropy.
 We find that the interplanar coupling introduced into the problem leads to a suppression
of the 2D quantum fluctuations caused by {\it intraplanar} anisotropies.
 For the large magnitude of $\Delta J_{\perp}=10^{-3}$, {\emph{i.e.}}
$\Delta J_{\perp}\sim 2\Delta\Gamma \ll d$, the net interplanar coupling $\Delta J_{\perp}$
dominates over the DM and XY-like pseudo-dipolar anisotropies (see
the lower curve in figure~\ref{Fig07}a).

 Finally, we conclude the presentation of these numerical results by returning to a discussion of the
correlation between the magnitude of the zone-centre spin-wave gaps and the
behaviour of the components $\chi^{x,z}$, $\chi^y$. One can see that only when
$\varepsilon_3$ is greater than $\varepsilon_2$ at zero interplanar coupling
$\Delta J_{\perp}=0$ does the $y$ component of the susceptibility at $T=0$ become
less than components $\chi^{x,z}$, since the ratio between gaps remains unchanged
for any values of $\Delta J_{\perp}$ (see equation~(\ref{eq:ratio_chi_Omega})).

\section{Approximate simple tetragonal model}

\begin{figure}[h]
\centerline{\epsfxsize .4\textwidth\epsfbox{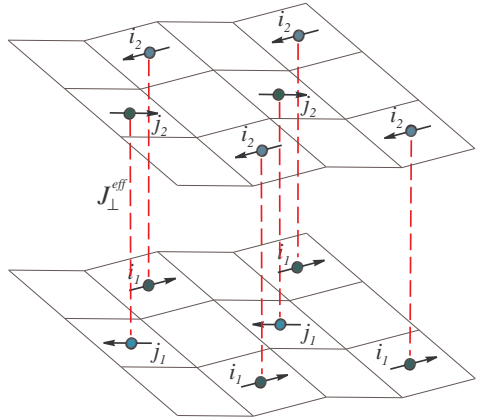}}
\caption{\label{fig:lattice2} (Colour online) Magnetic structure of the
         simple tetragonal lattice described by the effective
         model Hamiltonian of equation~(\ref{eq:H_cub}). Use of this
Hamiltonian for this lattice accurately approximates the magnetic
         response of the La$_2$CuO$_4$ crystal in the LTO phase.}
\end{figure}

 For the model parameters of interest our
initial Hamiltonian~(1) can be approximated by a simple tetragonal
model Hamiltonian which includes the intraplanar isotropic
Heisenberg interaction $J$, the anisotropic DM term ${\bf D}$ that alternates in
sign from bond to bond, the XY-like pseudo-dipolar anisotropy $\Delta\Gamma$, and an
effective interplanar interaction $J^{eff}_{\perp}$. This effective model
is defined by
\bea
\label{eq:H_cub} \fl   H&=&
       \sum_{\langle i,j\rangle}[J{\bf S}_{i}\cdot{\bf S}_{j}
       -\Delta\Gamma S^z_iS^z_j
       +{\bf D}_{ij}\cdot({\bf S}_{i}\times{\bf S}_{j})]
       +\sum_{\langle k,k'\rangle}J^{eff}_{\perp}{\bf S}_{k}\cdot{\bf S}_{k'}~~.
\eea
The single-plane effective Hamiltonian was proposed by Peters {\em et al.} \cite{Peters}
long ago, and its reliability was demonstrated in our previous work \cite{Tabunshchyk}.
Here the interplanar coupling $J^{eff}_{\perp}$ is added phenomenologically for
a simple tetragonal lattice (see figure~\ref{fig:lattice2}), $i$ and $j$ are
nearest-neighbour sites in the same CuO plane (indexes $i_1,j_1$ and $i_2,j_2$
in figure~\ref{fig:lattice2}) and $k$, $k'$ are nearest-neighbour
sites in adjacent planes (indexes $i_1,i_2$ and $j_1,j_2$
in figure~\ref{fig:lattice2}).
 Since the interplanar coordination number for interplanar interaction within a
simple tetragonal model is half that of the corresponding one for the
coupling $\Delta J_{\perp}$,
we can approximate $J^{eff}_{\perp}=2\Delta J_{\perp}$.

\begin{figure}[h]
\begin{center}
 \epsfxsize 0.47\textwidth\epsfbox{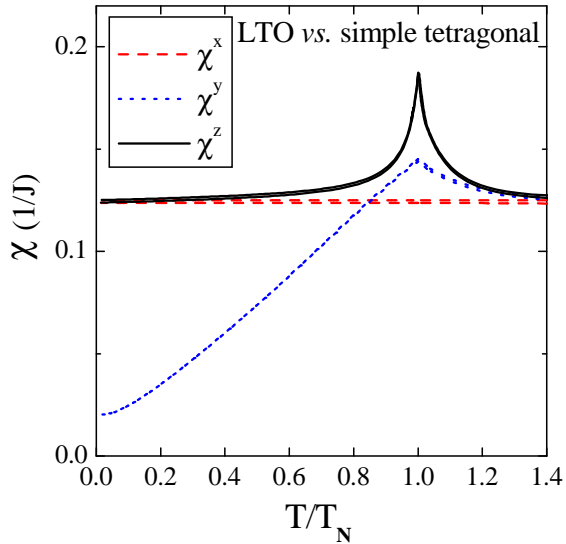}
\end{center}
\caption{\label{Fig04}(Colour online) The susceptibility, in units
of $1/J$, as a function of $T/T_N$. We present both
results from the RPA calculation on the initial model Hamiltonian of
equation~(1) with
$\Delta J_{\perp}/J=0.1\times 10^{-3}$ and the simple tetragonal model
with the Hamiltonian of
equation~(\ref{eq:H_cub}) with $J^{eff}_{\perp}/J=0.2\times 10^{-3}$
(for $d/J=0.02$ and $\Delta\Gamma/J=0.42\times10^{-3}$). The curves from these
two models essentially coincide, and no differences can seen on this scale. }
\end{figure}

 We performed the calculations of the order parameter, spin-wave excitations and
susceptibility within the RPA scheme for the effective simple tetragonal model
of equation~(\ref{eq:H_cub}),
and found that the transition temperature, spectrum, and behaviour of the order parameter
and susceptibility almost identical to that on the initial model of equation~(1).
 In figure~\ref{Fig04} we show representative data for the susceptibility obtained
for the initial body-centred orthorhombic model as well as for the effective simple tetragonal one
with $J^{eff}_{\perp}=2\Delta J_{\perp}$ -- clearly, the agreement between the predictions
for these two models is excellent, so when studying the magnetic properties of the model~(1) on
3D body-centred orthorhombic lattice one can utilize the effective Hamiltonian of the simple
tetragonal lattice.
 Consequently, the magnetism of the La$_2$CuO$_4$ system in the LTO phase can be modelled by the
Hamiltonian of equation~(\ref{eq:H_cub}).

\section{Conclusions and Discussion}

 In this paper we presented a theoretical investigation of the
body-centred orthorhombic lattice Heisenberg antiferromagnet with in-plane
symmetric and anti-symmetric anisotropies, and a weak anisotropic AFM interlayer coupling.
 Our study was focused on the role of the different interactions in explaining
the magnetic properties of a \LCO~crystal in the low-temperature orthorhombic
(LTO) phase.
 Due to the transition into the orthorhombic phase, the AFM interplanar coupling
between nearest-neighbour spins in the adjacent \CuO~planes exhibits a small
anisotropy.
 We have found that such a small anisotropy plays an important role in magnetic properties
of the system.
 In figures~\ref{Fig06}a,~\ref{Fig07}a one sees a significant change in the behaviour of magnetic
susceptibility as a function of temperature by varying the magnitude of the net interplanar coupling
$\Delta J_{\perp}$.
 We also obtained that the (larger) individual superexchange interaction between any two
nearest-neighbour spins in the adjacent planes does not affect the physics of the
model (figure~\ref{Fig01}).

 Our results have shown that in the case of an isotropic interplanar coupling,
2D quantum fluctuations dominate over the effects of the 3D interaction, and the transition to the long-range
magnetically ordered state, as well as the behaviour of the susceptibility, order parameter and magnon excitation
spectrum, are not influenced by the interplanar exchange coupling (however, for a 3D model the $z$ component of the susceptibility will not diverge, as it does in a 2D model).
 Thus, in the case of the body-centred lattice model of equation~(1) with an isotropic interplanar coupling
($J'_{\perp}=J_{\perp}$) one can analyze the system using a 2D square lattice model with
intraplanar anisotropies only.

 We have also obtained that the initial model Hamiltonian (1) can be effectively replaced by a simpler
one with fewer model parameters, namely by the AFM Heisenberg Hamiltonian with DM interaction,
XY-like pseudo-dipolar anisotropy, and an effective interplanar interaction $J^{eff}_{\perp}$ (added phenomenologically
for a simple tetragonal lattice).
 Here $J^{eff}_{\perp}/2\sim 10^{-4}$ describes the small anisotropy of the AFM interplanar coupling
in the initial system.

 We emphasize an important conclusion that can be drawn from our results.
 We have found that in-plane anisotropy introduced into the problem by symmetric XY-like pseudo-dipolar
and antisymmetric DM interactions largely determines the behaviour of the magnetic susceptibility,
transition temperature into the long-range order state, and the spin-wave gaps in the case of a
\emph{3D model} (within the wide range of model parameters of interest). Further,
even when one studies a 3D model,
 the effect of quantum fluctuations is very strong in all temperature
regions below the transition temperature, and cannot be ignored.
 Similar to the results of our previous paper \cite{Tabunshchyk}, we have also obtained the large short-range
correlations in the broad temperature region above the N\'eel temperature.

 Now we comment on the comparison of our results to the experimentally observed anisotropies
of the susceptibility \cite{Lavrov} and spin-wave gaps \cite{Keimer} that motivated our work.
 We can state that \emph{all anisotropic interactions} involved in the model, \emph{i.e.} DM,
XY-like pseudo-dipolar, and interplanar ones, are responsible for the unusual anisotropy in the
magnetic susceptibility, and the appearance of gaps in spin-wave excitation spectrum.
 More concretely, by comparing to a purely 2D model, the inclusion of interplanar anisotropy leads to
the splitting of either of the in- and out-of plane zone-centre spin-wave modes.
 While the neutron-scattering experiments find only two gaps, one in-plane mode $\varepsilon_i\approx2.3$~meV
and one out-of plane mode $\varepsilon_o\approx 5$~meV, we can infer the following possible situation
that is predicted from our results: the in-plane mode $\varepsilon_1$ (which is always larger than $\varepsilon_2$)
has a gap with a magnitude of about 10~meV. Indeed, such an in-plane gap can be seen from the result for the
spin-wave spectra in the neutron-scattering experiments \cite{Keimer}; other observed gaps corresponds to the
out-of plane mode $\varepsilon_3\approx 5$~meV and in-plane mode $\varepsilon_2\approx 2.3$~meV.
 The magnitude of the gap of the remaining out-of plane mode, $\varepsilon_4$, is relatively small
and apparently has not be seen by experiment.
 Therefore the following hierarchy $\varepsilon_1 > \varepsilon_3 > \varepsilon_2 > \varepsilon_4$ agrees
with the experiment.
 In this paper we established the correlation between the ratio of the in and out-of-plane gaps of the excitation
spectrum and the behaviour of $\chi^{x,z}$ and $\chi^{y}$ components of susceptibility in the zero
temperature limit.
 However, the proposed hierarchy of the spin-wave gaps takes place only if the ratio between the $x$ and $y$
components is opposite to that observed in experiment ($\chi^x<\chi^y$).
 This necessarily leads to the question, would other interactions, \textit {e.g.} ring exchange and/or
the interaction between the next nearest neighbour sites \cite{Katanin}, lead to an accurate explanation of the
susceptibility data within the RPA scheme?

 In order to answer this question we have performed calculation for the square lattice AFM Heisenberg
model with the DM and XY-like pseudo-dipolar anisotropies by additionally taking into account
the ring exchange and the interactions between the next nearest neighbour sites (for the energy
scales of these additional interactions see reference \cite{Katanin}).
 Our results of the RPA calculations have established that ring exchange together with the second and third nearest
neighbour in-plane exchanges {\emph{do not change}} the results presented in our
earlier paper \cite{Tabunshchyk} regarding the correlation between the ratio of the
in and out-of-plane spin-wave modes gaps and the behaviour of the
$\chi^x$ and $\chi^y$ components of susceptibility in the zero-$T$
limit, {\emph{viz.}}
$\varepsilon_o^2/\varepsilon_i^2\approx\chi^x/\chi^y$. So, physics beyond
what has been presented in our previous and this manuscript is important, but that does not
imply that a more complicated Hamiltonian with more interactions is necessarily required.

A potential resolution of this dilemma can be found in studies based on the quantum non-linear
sigma model \cite{Neto}. However, within a theory that accounts for short-wavelength behaviour
we will show in a future publication how the ``next" approximation beyond that used in our
previous \cite{Tabunshchyk} and present papers fits the experimental data. This allows for the
important next problem of the coupling of the anisotropic AFM state to either
localized or mobile holes to be examined.

\ack

We thank Alexander Lavrov, Yoichi Ando, and Marcello Barbosa da Silva Neto for helpful comments.
This work was supported by the NSERC of Canada, and NATO.

\newpage

\appendix

\section{Characteristic representation}

 The transformation between the initial representation and the characteristic representation (CR)
in which the quantization axis $\sigma^z$ is in the direction of the classical moment
(see figure~1) reads
\bea
\label{eq:1_rot1}
\left (
 \begin{array}{c}
        \sigma_{i_1}^x\\
        \sigma_{i_1}^y\\
        \sigma_{i_1}^z\\
 \end{array}
\right )
&=&\frac 1{\sqrt{2}}
\left (
 \begin{array}{ccc}
   1 &\sin\theta &-\cos\theta\\
  -1 &\sin\theta &-\cos\theta\\
  0&\sqrt{2}\cos\theta &\sqrt{2}\sin\theta\\
  \end{array}
 \right )
\left (
 \begin{array}{c}
        S_{i_1}^x\\
        S_{i_1}^y\\
        S_{i_1}^z\\
 \end{array}
\right ),\\
\label{eq:21_rot}
\left (
 \begin{array}{c}
        \sigma_{j_1}^x\\
        \sigma_{j_1}^y\\
        \sigma_{j_1}^z\\
 \end{array}
\right )&=&\frac 1{\sqrt{2}}
\left (
 \begin{array}{ccc}
   1 &\sin\theta &\cos\theta\\
  -1 &\sin\theta &\cos\theta\\
  0&-\sqrt{2}\cos\theta &\sqrt{2}\sin\theta\\
  \end{array}
 \right )
\left (
 \begin{array}{c}
        S_{j_1}^x\\
        S_{j_1}^y\\
        S_{j_1}^z\\
 \end{array}
\right ),\\
\label{eq:1_rot3}
\left (
 \begin{array}{c}
        \sigma_{i_2}^x\\
        \sigma_{i_2}^y\\
        \sigma_{i_2}^z\\
 \end{array}
\right )
&=&\frac 1{\sqrt{2}}
\left (
 \begin{array}{ccc}
   1 &-\sin\theta &\cos\theta\\
  -1 &-\sin\theta &\cos\theta\\
  0&-\sqrt{2}\cos\theta &-\sqrt{2}\sin\theta\\
  \end{array}
 \right )
\left (
 \begin{array}{c}
        S_{i_2}^x\\
        S_{i_2}^y\\
        S_{i_2}^z\\
 \end{array}
\right ),\\
\label{eq:1_rot4}
\left (
 \begin{array}{c}
        \sigma_{j_2}^x\\
        \sigma_{j_2}^y\\
        \sigma_{j_2}^z\\
 \end{array}
\right )&=&\frac 1{\sqrt{2}}
\left (
 \begin{array}{ccc}
   1 &-\sin\theta &-\cos\theta\\
  -1 &-\sin\theta &-\cos\theta\\
  0&\sqrt{2}\cos\theta &-\sqrt{2}\sin\theta\\
  \end{array}
 \right )
\left (
 \begin{array}{c}
        S_{j_2}^x\\
        S_{j_2}^y\\
        S_{j_2}^z\\
 \end{array}
\right ).
\eea

 In our calculations we formally divided the body-centred lattice into the four sublattices
which differ in orientation of their spins within the ground state.
 As was explained in section~\ref{sec:Model}, we follow the notation where $i_1$-sites belong to sublattice~1,
$j_1$-sites to sublattice~2, $i_2$-sites to sublattice~3, and $j_2$-sites to sublattice~4, respectively.

 By using the CR as the basic representation we obtain the components of susceptibility
that determine the response of the expectation value of the spins in one sublattice with
respect to the external magnetic field applied to another one.
 For instance, the component
\bea
\label{eq:iterpret1}
\chi^{\sigma^x\sigma^y}_{13}\equiv
\frac 4N\frac {\partial}{\partial h_3^{y}}\sum^{N/4}_{i_1=1}\langle\sigma_{i_1}^{x}\rangle=
\frac 4N\sum^{N/4}_{i_1=1}\sum^{N/4}_{i_2=1}
 \int^{\beta}_0\!\!\langle T_{\tau}\sigma_{i_1}^{x}(\tau)
 \sigma_{i_2}^{y}(0)\rangle{\rm d}\tau~~,
\eea
determines the response of the expectation value of the spin $\sigma^{x}$,
$4/N\sum^{N/4}_{i_1}\langle\sigma_{i_1}^{x}\rangle$, in
sublattice '1' to the field applied to sublattice '3' in the
$\sigma^y$-direction of the corresponding coordinate system within
the CR \cite{Tabunshchyk}.

 The transformation between the susceptibility components in the initial and CR coordinates reads
\bea
\nonumber
\fl\chi^x&=&\frac 12\{
   \chi^{\sigma^x\sigma^x}_{11}+\chi^{\sigma^x\sigma^x}_{12}+\chi^{\sigma^x\sigma^x}_{13}+\chi^{\sigma^x\sigma^x}_{14}
  -\chi^{\sigma^x\sigma^y}_{11}-\chi^{\sigma^x\sigma^y}_{12}-\chi^{\sigma^x\sigma^y}_{13}-\chi^{\sigma^x\sigma^y}_{14}\\
\label{eq:CRtoINx}
\fl&&+\chi^{\sigma^y\sigma^y}_{11}+\chi^{\sigma^y\sigma^y}_{12}+\chi^{\sigma^y\sigma^y}_{13}+\chi^{\sigma^y\sigma^y}_{14}
  -\chi^{\sigma^y\sigma^x}_{11}-\chi^{\sigma^y\sigma^x}_{12}-\chi^{\sigma^y\sigma^x}_{13}-\chi^{\sigma^y\sigma^x}_{14}
   \},\\
\nonumber
\fl\chi^y&=&\frac{\sin^2(\theta)}2\{
   \chi^{\sigma^x\sigma^x}_{11}+\chi^{\sigma^x\sigma^x}_{12}-\chi^{\sigma^x\sigma^x}_{13}-\chi^{\sigma^x\sigma^x}_{14}
  +\chi^{\sigma^x\sigma^y}_{11}+\chi^{\sigma^x\sigma^y}_{12}-\chi^{\sigma^x\sigma^y}_{13}-\chi^{\sigma^x\sigma^y}_{14}\\
\nonumber
\fl&&\qquad
  +\chi^{\sigma^y\sigma^y}_{11}+\chi^{\sigma^y\sigma^y}_{12}-\chi^{\sigma^y\sigma^y}_{13}-\chi^{\sigma^y\sigma^y}_{14}
  +\chi^{\sigma^y\sigma^x}_{11}+\chi^{\sigma^y\sigma^x}_{12}-\chi^{\sigma^y\sigma^x}_{13}-\chi^{\sigma^y\sigma^x}_{14}
   \}\\
\label{eq:CRtoINy}
\fl&+&\cos^2(\theta)\{
   \chi^{\sigma^z\sigma^z}_{11}-\chi^{\sigma^z\sigma^z}_{12}-\chi^{\sigma^z\sigma^z}_{13}+\chi^{\sigma^z\sigma^z}_{14}
   \},\\
\nonumber
\fl\chi^z&=&\frac{\cos^2(\theta)}2\{
   \chi^{\sigma^x\sigma^x}_{11}-\chi^{\sigma^x\sigma^x}_{12}-\chi^{\sigma^x\sigma^x}_{13}+\chi^{\sigma^x\sigma^x}_{14}
  +\chi^{\sigma^x\sigma^y}_{11}-\chi^{\sigma^x\sigma^y}_{12}-\chi^{\sigma^x\sigma^y}_{13}+\chi^{\sigma^x\sigma^y}_{14}\\
\nonumber
\fl&&\qquad
  +\chi^{\sigma^y\sigma^y}_{11}-\chi^{\sigma^y\sigma^y}_{12}-\chi^{\sigma^y\sigma^y}_{13}+\chi^{\sigma^y\sigma^y}_{14}
  +\chi^{\sigma^y\sigma^x}_{11}-\chi^{\sigma^y\sigma^x}_{12}-\chi^{\sigma^y\sigma^x}_{13}+\chi^{\sigma^y\sigma^x}_{14}
   \}\\
\label{eq:CRtoINz}
\fl&+&\sin^2(\theta)\{
   \chi^{\sigma^z\sigma^z}_{11}+\chi^{\sigma^z\sigma^z}_{12}-\chi^{\sigma^z\sigma^z}_{13}-\chi^{\sigma^z\sigma^z}_{14}
   \}.
\eea

\section{System of equations for the Green's functions in the momentum-frequency representation}

 Let us present the most general form of the system of equations for the Green's function's
in equation~(\ref{eq:GGG}).
 We introduce notation $g$ for the zero-order $g_{ln}=G_{ln}(\bk,\omega_m)$ and/or first-order
$g_{ln}=G^{(1)}_{ln}(\bk,\omega_m)$ Green's functions depending on the coefficients in the equations (see below).
 Then, the system reads
\bea
\nonumber
\fl\frac{{\ri}\omega_m}{2{\cal Z}\eta} g_{12} &=& \hphantom{-}U_{12} +
                   \frac{J_{mfa}}2g_{12}+A_{\bk}g_{22}-B_{\bk}g^-_{22}
                          +a_{\bk}g_{42}-{\ri}b_{\bk}g^-_{42}-{\ri}d_{\bk}g^-_{32}~~,\\
\nonumber
\fl\frac{{\ri}\omega_m}{2{\cal Z}\eta} g_{22} &=& \hphantom{-}U_{22} +
                   \frac{J_{mfa}}2g_{22}+A_{\bk}g_{12}-B_{\bk}g^-_{12}
                          +a_{\bk}g_{32}-{\ri}b_{\bk}g^-_{32}-{\ri}d_{\bk}g^-_{42}~~,\\
\nonumber
\fl\frac{{\ri}\omega_m}{2{\cal Z}\eta} g^-_{12} &=& -U^-_{12} -
                   \frac{J_{mfa}}2g^-_{12}-A_{\bk}g^-_{22}+B^*_{\bk}g_{22}
                            -a_{\bk}g^-_{42}-{\ri}b_{\bk}g_{42}-{\ri}d_{\bk}g_{32}~~,\\
\label{eq:system_g}
\fl\frac{{\ri}\omega_m}{2{\cal Z}\eta} g^-_{22} &=& -U^-_{22} -
                   \frac{J_{mfa}}2g^-_{22}-A_{\bk}g^-_{12}+B^*_{\bk}g_{12}
                            -a_{\bk}g^-_{32}-{\ri}b_{\bk}g_{32}-{\ri}d_{\bk}g_{42}~~,\\
\nonumber
\fl\frac{{\ri}\omega_m}{2{\cal Z}\eta} g_{32} &=& \hphantom{-}U_{32} +
                   \frac{J_{mfa}}2g_{32}+A_{\bk}g_{42}+B^*_{\bk}g^-_{42}
                          +a_{\bk}g_{22}-{\ri}b_{\bk}g^-_{22}-{\ri}d_{\bk}g^-_{12}~~,\\
\nonumber
\fl\frac{{\ri}\omega_m}{2{\cal Z}\eta} g_{42} &=& \hphantom{-}U_{42} +
                   \frac{J_{mfa}}2g_{42}+A_{\bk}g_{32}+B^*_{\bk}g^-_{32}
                          +a_{\bk}g_{12}-{\ri}b_{\bk}g^-_{12}-{\ri}d_{\bk}g^-_{22}~~,\\
\nonumber
\fl\frac{{\ri}\omega_m}{2{\cal Z}\eta} g^-_{32} &=& -U^-_{32} -
                   \frac{J_{mfa}}2g^-_{32}-A_{\bk}g^-_{42}-B_{\bk}g_{42}
                            -a_{\bk}g^-_{22}-{\ri}b_{\bk}g_{22}-{\ri}d_{\bk}g_{12}~~,\\
\nonumber
\fl\frac{{\ri}\omega_m}{2{\cal Z}\eta} g^-_{42} &=& -U^-_{42} -
                   \frac{J_{mfa}}2g^-_{42}-A_{\bk}g^-_{32}-B_{\bk}g_{32}
                            -a_{\bk}g^-_{12}-{\ri}b_{\bk}g_{12}-{\ri}d_{\bk}g_{22}~~.
\eea
 In the case of the zero-order system, $g_{ln}=G_{ln}(\bk,\omega_m)$, we have
\bea
\label{coef_G0}
\fl U_{22}=-\frac 1{\cal Z},\quad U_{12}=U^-_{12}=U^-_{22}=U_{32}=U_{42}=U^-_{32}=U^-_{42}=0.
\eea
  In case of the first-order system $g_{ln}=G^{(1)}_{ln}(\bk,\omega_m)$ we have
\bea
\nonumber
&&\fl
U_{12} = V_{12}G_{12} =
 \bigg\{     \frac 1{2{\cal Z}\eta}{+} \frac{J_2}2 \frac{\rv_2}{\eta}
                                  +\frac{J_\perp}2 \frac{\rv_3}{\eta}
                                      -\frac{J_p}2 \frac{\rv_4}{\eta}
                                 +\bigg(\frac{{\ri}\omega}{2{\cal Z}\eta}{-}\frac{J_{mfa}}2\bigg)\frac{\rv_1}{\eta}\bigg\}G_{12},\\
\nonumber
&&\fl
U_{22} =V_{22}G_{22} =
 \bigg\{ \hphantom{\frac 1{2{\cal Z}\eta}+}  \frac{J_2}2 \frac{\rv_1}{\eta}
                                  +\frac{J_\perp}2 \frac{\rv_4}{\eta}
                                      -\frac{J_p}2 \frac{\rv_3}{\eta}
                                 +\bigg(\frac{{\ri}\omega}{2{\cal Z}\eta}{-}\frac{J_{mfa}}2\bigg)\frac{\rv_2}{\eta}\bigg\}G_{22},\\
\nonumber
&&\fl
U^-_{12} =V^-_{12} G^-_{12} =
 \bigg\{   \frac 1{2{\cal Z}\eta}{+} \frac{J_2}2 \frac{\rv_2}{\eta}
                                  +\frac{J_\perp}2 \frac{\rv_3}{\eta}
                                      -\frac{J_p}2 \frac{\rv_4}{\eta}
                                 -\bigg(\frac{{\ri}\omega}{2{\cal Z}\eta}{+}\frac{J_{mfa}}2\bigg)\frac{\rv_1}{\eta}\bigg\}G^-_{12},\\
\label{eq:coef_G1}
&&\fl
U^-_{22} =V^-_{22} G^-_{22} =
 \bigg\{ \hphantom{\frac 1{2{\cal Z}\eta}+} \frac{J_2}2 \frac{\rv_1}{\eta}
                                  +\frac{J_\perp}2 \frac{\rv_4}{\eta}
                                      -\frac{J_p}2 \frac{\rv_3}{\eta}
                                 -\bigg(\frac{{\ri}\omega}{2{\cal Z}\eta}{+}\frac{J_{mfa}}2\bigg)\frac{\rv_2}{\eta}\bigg\}G^-_{22},\\
\nonumber
&&\fl
U_{32} =V_{32} G_{32} =
\bigg\{ \hphantom{\frac 1{2{\cal Z}\eta}+}   \frac{J_2}2 \frac{\rv_4}{\eta}
                                  +\frac{J_\perp}2 \frac{\rv_1}{\eta}
                                      -\frac{J_p}2 \frac{\rv_2}{\eta}
                                 +\bigg(\frac{{\ri}\omega}{2{\cal Z}\eta}{-}\frac{J_{mfa}}2\bigg)\frac{\rv_3}{\eta}\bigg\}G_{32},\\
\nonumber
&&\fl
U_{42} =V_{42} G_{42} =
\bigg\{ \hphantom{\frac 1{2{\cal Z}\eta}+}  \frac{J_2}2 \frac{\rv_3}{\eta}
                                  +\frac{J_\perp}2 \frac{\rv_2}{\eta}
                                      -\frac{J_p}2 \frac{\rv_1}{\eta}
                                 +\bigg(\frac{{\ri}\omega}{2{\cal Z}\eta}{-}\frac{J_{mfa}}2\bigg)\frac{\rv_4}{\eta}\bigg\}G_{42},\\
\nonumber
&&\fl
U^-_{32} =V^-_{32} G^-_{32} =
\bigg\{\hphantom{\frac 1{2{\cal Z}\eta}+} \frac{J_2}2 \frac{\rv_4}{\eta}
                                  +\frac{J_\perp}2 \frac{\rv_1}{\eta}
                                      -\frac{J_p}2 \frac{\rv_2}{\eta}
                                 -\bigg(\frac{{\ri}\omega}{2{\cal Z}\eta}{+}\frac{J_{mfa}}2\bigg)\frac{\rv_3}{\eta}\bigg\}G^-_{32},\\
\nonumber
&&\fl
U^-_{42} =V^-_{42} G^-_{42} =
\bigg\{\hphantom{\frac 1{2{\cal Z}\eta}+} \frac{J_2}2 \frac{\rv_3}{\eta}
                                  +\frac{J_\perp}2 \frac{\rv_2}{\eta}
                                      -\frac{J_p}2 \frac{\rv_1}{\eta}
                                 -\bigg(\frac{{\ri}\omega}{2{\cal Z}\eta}{+}\frac{J_{mfa}}2\bigg)\frac{\rv_4}{\eta}\bigg\}G^-_{42};
\eea
where the new quantities $V_{ln}$ and $V^{-}_{ln}$ were introduced.

\section{Zero-order Green's functions}
\label{eq:solutionG}
 For ease of presentation, we define some new quantities:
\bea
\nonumber
\fl && x_{1,\bk}=-2a_{\bk}[a_{\bk}(A_{\bk}{-}J_{mfa}/2)-(b_{\bk}{-}d_{\bk})\Im B_{\bk}],\qquad
       y_{1,\bk}=-(A_{\bk}{-}J_{mfa}/2),\\
\nonumber
\fl && x_{2,\bk}=\hphantom{-}
                  2a_{\bk}[a_{\bk}(A_{\bk}{+}J_{mfa}/2)-(b_{\bk}{+}d_{\bk})\Im B_{\bk}],\qquad
       y_{2,\bk}=\hphantom{-}(A_{\bk}{+}J_{mfa}/2),
\eea
\bea
\nonumber
\fl&& x_{3,\bk}=  2\Re B_{\bk}[-a^2_{\bk} + (b_{\bk}{-}d_{\bk})^2] +
             2{\ri}(b_{\bk}{-}d_{\bk})[\hphantom{-} a_{\bk}(A_{\bk}{-}J_{mfa}/2)-\Im B_{\bk}(b_{\bk}{-}d_{\bk})],\\
\nonumber
\fl&&
   x_{4,\bk}=     2\Re B_{\bk}[\hphantom{-} a^2_{\bk} - (b_{\bk}{+}d_{\bk})^2] +
             2{\ri}(b_{\bk}{+}d_{\bk})[-a_{\bk}(A_{\bk}{+}J_{mfa}/2)+\Im B_{\bk}(b_{\bk}{+}d_{\bk})],\\
\nonumber
\fl&&
   y_{3,\bk}=    -B^*_{\bk},\\
\nonumber
\fl&&
   y_{4,\bk}= \hphantom{-} B^*_{\bk},
\eea
\bea
\nonumber
\fl&& x_{5,\bk}=- 2(A_{\bk}{-}J_{mfa}/2)[a_{\bk}(A_{\bk}{-}J_{mfa}/2)-(b_{\bk}{-}d_{\bk})\Im B_{\bk}]
             + 2a_{\bk}\Re^2 B_{\bk}, \\
\nonumber
\fl && \hspace*{1cm}
+2{\ri}\Re B_{\bk}[(A_{\bk}{-}J_{mfa}/2)(b_{\bk}{-}d_{\bk}) - a_{\bk}\Im B_{\bk}],\\
\nonumber
\fl&& x_{6,\bk}=\hphantom{-}
                  2(A_{\bk}{+}J_{mfa}/2)[a_{\bk}(A_{\bk}{+}J_{mfa}/2)-(b_{\bk}{+}d_{\bk})\Im B_{\bk}]
             - 2a_{\bk}\Re^2 B_{\bk}, \\
\nonumber
\fl && \hspace*{1cm}
-2{\ri}\Re B_{\bk}[(A_{\bk}{+}J_{mfa}/2)(b_{\bk}{+}d_{\bk}) - a_{\bk}\Im B_{\bk}],\\
\nonumber
\fl&& y_{5,\bk}=          -  a_{\bk},\qquad
      z_{5,\bk}= - 2\Im B_{\bk}(b_{\bk}{-}d_{\bk}) + 2a_{\bk}(A_{\bk}{-}J_{mfa}/2)- 2{\ri}\Re B_{\bk}(b_{\bk}{-}d_{\bk}),\\
\nonumber
\fl&& y_{6,\bk}=\hphantom{-} a_{\bk},\qquad
      z_{6,\bk}= - 2\Im B_{\bk}(b_{\bk}{+}d_{\bk}) + 2a_{\bk}(A_{\bk}{+}J_{mfa}/2)- 2{\ri}\Re B_{\bk}(b_{\bk}{+}d_{\bk}),
\eea
\bea
\nonumber
\fl&& x_{7,\bk}= \hphantom{-}
         2{\ri} [ a_{\bk}(A_{\bk}{-}J_{mfa}/2)\Im B_{\bk}-(b_{\bk}{-}d_{\bk})|B_{\bk}|^2],\\
\nonumber
\fl&& x_{8,\bk}= -
         2{\ri} [ a_{\bk}(A_{\bk}{+}J_{mfa}/2)\Im B_{\bk}-(b_{\bk}{+}d_{\bk})|B_{\bk}|^2],\\
\nonumber
\fl&& y_{7,\bk}=\hphantom{-} {\ri}(b_{\bk}{-}d_{\bk}),\qquad  z_{7,\bk}= - 2a_{\bk}\Re B_{\bk},\\
\nonumber
\fl&& y_{8,\bk}=          -  {\ri}(b_{\bk}{+}d_{\bk}),\qquad  z_{8,\bk}= - 2a_{\bk}\Re B_{\bk}.
\eea
Then, the solution of the system of equations for the zero-order Green's functions can be written as
\bea
\fl G_{12}(\bk,\omega_m)=\frac{\eta}{4}\bigg\{\!\!&\hphantom{+}&
         \frac 1{\omega_{1,\bk}}\bigg(
                 \frac{\omega_{1,\bk}{+}y_{1,\bk}{+}x_{1,\bk}/\sqrt{\beta_{1,\bk}}}{{\ri}\omega_m-\varepsilon_{1,\bk}}
                +\frac{\omega_{1,\bk}{-}y_{1,\bk}{-}x_{1,\bk}/\sqrt{\beta_{1,\bk}}}{{\ri}\omega_m+\varepsilon_{1,\bk}}
           \bigg)\\
\nonumber
\fl\!\!&+& \frac 1{\omega_{2,\bk}}\bigg(
                 \frac{\omega_{2,\bk}{+}y_{1,\bk}{-}x_{1,\bk}/\sqrt{\beta_{1,\bk}}}{{\ri}\omega_m-\varepsilon_{2,\bk}}
                +\frac{\omega_{2,\bk}{-}y_{1,\bk}{+}x_{1,\bk}/\sqrt{\beta_{1,\bk}}}{{\ri}\omega_m+\varepsilon_{2,\bk}}
           \bigg)\\
\nonumber
\fl\!\!&-& \frac 1{\omega_{3,\bk}}\bigg(
                 \frac{\omega_{3,\bk}{+}y_{2,\bk}{+}x_{2,\bk}/\sqrt{\beta_{2,\bk}}}{{\ri}\omega_m-\varepsilon_{3,\bk}}
                +\frac{\omega_{3,\bk}{-}y_{2,\bk}{-}x_{2,\bk}/\sqrt{\beta_{2,\bk}}}{{\ri}\omega_m+\varepsilon_{3,\bk}}
           \bigg)\\
\nonumber
\fl\!\!&-& \frac 1{\omega_{4,\bk}}\bigg(
                 \frac{\omega_{4,\bk}{+}y_{2,\bk}{-}x_{2,\bk}/\sqrt{\beta_{2,\bk}}}{{\ri}\omega_m-\varepsilon_{4,\bk}}
                +\frac{\omega_{4,\bk}{-}y_{2,\bk}{+}x_{2,\bk}/\sqrt{\beta_{2,\bk}}}{{\ri}\omega_m+\varepsilon_{4,\bk}}
           \bigg) \bigg\}~~,\\
\label{eq:G22zero}
\fl G_{22}(\bk,\omega_m)=\frac{\eta}{4}\bigg\{\!\!&-&
         \frac 1{\omega_{1,\bk}}\bigg(
           \frac{\omega_{1,\bk}{+}y_{1,\bk}{+}x_{1,\bk}/\sqrt{\beta_{1,\bk}}}{{\ri}\omega_m-\varepsilon_{1,\bk}}
          +\frac{\omega_{1,\bk}{-}y_{1,\bk}{-}x_{1,\bk}/\sqrt{\beta_{1,\bk}}}{{\ri}\omega_m+\varepsilon_{1,\bk}}
           \bigg)\\
\nonumber
\fl\!\!&-& \frac 1{\omega_{2,\bk}}\bigg(
                 \frac{\omega_{2,\bk}{+}y_{1,\bk}{-}x_{1,\bk}/\sqrt{\beta_{1,\bk}}}{{\ri}\omega_m-\varepsilon_{2,\bk}}
                +\frac{\omega_{2,\bk}{-}y_{1,\bk}{+}x_{1,\bk}/\sqrt{\beta_{1,\bk}}}{{\ri}\omega_m+\varepsilon_{2,\bk}}
           \bigg)\\
\nonumber
\fl\!\!&-& \frac 1{\omega_{3,\bk}}\bigg(
                 \frac{\omega_{3,\bk}{+}y_{2,\bk}{+}x_{2,\bk}/\sqrt{\beta_{2,\bk}}}{{\ri}\omega_m-\varepsilon_{3,\bk}}
                +\frac{\omega_{3,\bk}{-}y_{2,\bk}{-}x_{2,\bk}/\sqrt{\beta_{2,\bk}}}{{\ri}\omega_m+\varepsilon_{3,\bk}}
           \bigg)\\
\nonumber
\fl\!\!&-& \frac 1{\omega_{4,\bk}}\bigg(
                 \frac{\omega_{4,\bk}{+}y_{2,\bk}{-}x_{2,\bk}/\sqrt{\beta_{2,\bk}}}{{\ri}\omega_m-\varepsilon_{4,\bk}}
                +\frac{\omega_{4,\bk}{-}y_{2,\bk}{+}x_{2,\bk}/\sqrt{\beta_{2,\bk}}}{{\ri}\omega_m+\varepsilon_{4,\bk}}
           \bigg)
\bigg\}
\eea
\bea
\fl G^-_{12}(\bk,\omega_m)=\frac{\eta}{4}\bigg\{\!\!&\hphantom{+}&
         \frac 1{\omega_{1,\bk}}\bigg(
                 \frac{ y_{3,\bk}{+}x_{3,\bk}/\sqrt{\beta_{1,\bk}}}{{\ri}\omega_m-\varepsilon_{1,\bk}}
                +\frac{-y_{3,\bk}{-}x_{3,\bk}/\sqrt{\beta_{1,\bk}}}{{\ri}\omega_m+\varepsilon_{1,\bk}}
           \bigg)\\
\nonumber
\fl\!\!&+& \frac 1{\omega_{2,\bk}}\bigg(
                 \frac{ y_{3,\bk}{-}x_{3,\bk}/\sqrt{\beta_{1,\bk}}}{{\ri}\omega_m-\varepsilon_{2,\bk}}
                +\frac{-y_{3,\bk}{+}x_{3,\bk}/\sqrt{\beta_{1,\bk}}}{{\ri}\omega_m+\varepsilon_{2,\bk}}
           \bigg)\\
\nonumber
\fl\!\!&-& \frac 1{\omega_{3,\bk}}\bigg(
                 \frac{ y_{4,\bk}{+}x_{4,\bk}/\sqrt{\beta_{2,\bk}}}{{\ri}\omega_m-\varepsilon_{3,\bk}}
                +\frac{-y_{4,\bk}{-}x_{4,\bk}/\sqrt{\beta_{2,\bk}}}{{\ri}\omega_m+\varepsilon_{3,\bk}}
           \bigg)\\
\nonumber
\fl\!\!&-& \frac 1{\omega_{4,\bk}}\bigg(
                 \frac{ y_{4,\bk}{-}x_{4,\bk}/\sqrt{\beta_{2,\bk}}}{{\ri}\omega_m-\varepsilon_{4,\bk}}
                +\frac{-y_{4,\bk}{+}x_{4,\bk}/\sqrt{\beta_{2,\bk}}}{{\ri}\omega_m+\varepsilon_{4,\bk}}
           \bigg)
\bigg\}\\
\fl G^-_{22}(\bk,\omega_m)=\frac{\eta}{4}\bigg\{\!\!&-&
         \frac 1{\omega_{1,\bk}}\bigg(
                \frac{ y_{3,\bk}{+}x_{3,\bk}/\sqrt{\beta_{1,\bk}}}{{\ri}\omega_m-\varepsilon_{1,\bk}}
               +\frac{-y_{3,\bk}{-}x_{3,\bk}/\sqrt{\beta_{1,\bk}}}{{\ri}\omega_m+\varepsilon_{1,\bk}}
           \bigg)\\
\nonumber
\fl\!\!&-& \frac 1{\omega_{2,\bk}}\bigg(
                 \frac{ y_{3,\bk}{-}x_{3,\bk}/\sqrt{\beta_{1,\bk}}}{{\ri}\omega_m-\varepsilon_{2,\bk}}
                +\frac{-y_{3,\bk}{+}x_{3,\bk}/\sqrt{\beta_{1,\bk}}}{{\ri}\omega_m+\varepsilon_{2,\bk}}
           \bigg)\\
\nonumber
\fl\!\!&-& \frac 1{\omega_{3,\bk}}\bigg(
                 \frac{ y_{4,\bk}{+}x_{4,\bk}/\sqrt{\beta_{2,\bk}}}{{\ri}\omega_m-\varepsilon_{3,\bk}}
                +\frac{-y_{4,\bk}{-}x_{4,\bk}/\sqrt{\beta_{2,\bk}}}{{\ri}\omega_m+\varepsilon_{3,\bk}}
           \bigg)\\
\nonumber
\fl\!\!&-& \frac 1{\omega_{4,\bk}}\bigg(
                 \frac{ y_{4,\bk}{-}x_{4,\bk}/\sqrt{\beta_{2,\bk}}}{{\ri}\omega_m-\varepsilon_{4,\bk}}
                +\frac{-y_{4,\bk}{+}x_{4,\bk}/\sqrt{\beta_{2,\bk}}}{{\ri}\omega_m+\varepsilon_{4,\bk}}
           \bigg)
\bigg\}
\eea
\bea
\nonumber
\fl G_{32}(\bk,\omega_m)=\frac{\eta}{4}\bigg\{\!\!&\hphantom{+}&
         \frac 1{\omega_{1,\bk}}\bigg(
                    \frac{y_{5,\bk}{+}(x_{5,\bk}{+}z_{5,\bk}\omega_{1,\bk})/\sqrt{\beta_{1,\bk}}}{{\ri}\omega_m-\varepsilon_{1,\bk}}
                +\frac{{-}y_{5,\bk}{-}(x_{5,\bk}{-}z_{5,\bk}\omega_{1,\bk})/\sqrt{\beta_{1,\bk}}}{{\ri}\omega_m+\varepsilon_{1,\bk}}
           \bigg)\\
\nonumber
\fl\!\!&+& \frac 1{\omega_{2,\bk}}\bigg(
                    \frac{y_{5,\bk}{-}(x_{5,\bk}{+}z_{5,\bk}\omega_{2,\bk})/\sqrt{\beta_{1,\bk}}}{{\ri}\omega_m-\varepsilon_{2,\bk}}
                +\frac{{-}y_{5,\bk}{+}(x_{5,\bk}{-}z_{5,\bk}\omega_{2,\bk})/\sqrt{\beta_{1,\bk}}}{{\ri}\omega_m+\varepsilon_{2,\bk}}
           \bigg)\\
\nonumber
\fl\!\!&-& \frac 1{\omega_{3,\bk}}\bigg(
                    \frac{y_{6,\bk}{+}(x_{6,\bk}{+}z_{6,\bk}\omega_{3,\bk})/\sqrt{\beta_{2,\bk}}}{{\ri}\omega_m-\varepsilon_{3,\bk}}
                +\frac{{-}y_{6,\bk}{-}(x_{6,\bk}{-}z_{6,\bk}\omega_{3,\bk})/\sqrt{\beta_{2,\bk}}}{{\ri}\omega_m+\varepsilon_{3,\bk}}
           \bigg)\\
\nonumber
\fl\!\!&-& \frac 1{\omega_{4,\bk}}\bigg(
                    \frac{y_{6,\bk}{-}(x_{6,\bk}{+}z_{6,\bk}\omega_{4,\bk})/\sqrt{\beta_{2,\bk}}}{{\ri}\omega_m-\varepsilon_{4,\bk}}
                +\frac{{-}y_{6,\bk}{+}(x_{6,\bk}{-}z_{6,\bk}\omega_{4,\bk})/\sqrt{\beta_{2,\bk}}}{{\ri}\omega_m+\varepsilon_{4,\bk}}
           \bigg)
\bigg\}\\
\fl\\
\nonumber
\fl G_{42}(\bk,\omega_m)=\frac{\eta}{4}\bigg\{\!\!&-&
         \frac 1{\omega_{1,\bk}}\bigg(
                    \frac{y_{5,\bk}{+}(x_{5,\bk}{+}z_{5,\bk}\omega_{1,\bk})/\sqrt{\beta_{1,\bk}}}{{\ri}\omega_m-\varepsilon_{1,\bk}}
                +\frac{{-}y_{5,\bk}{-}(x_{5,\bk}{-}z_{5,\bk}\omega_{1,\bk})/\sqrt{\beta_{1,\bk}}}{{\ri}\omega_m+\varepsilon_{1,\bk}}
           \bigg)\\
\nonumber
\fl\!\!&-& \frac 1{\omega_{2,\bk}}\bigg(
                    \frac{y_{5,\bk}{-}(x_{5,\bk}{+}z_{5,\bk}\omega_{2,\bk})/\sqrt{\beta_{1,\bk}}}{{\ri}\omega_m-\varepsilon_{2,\bk}}
                +\frac{{-}y_{5,\bk}{+}(x_{5,\bk}{-}z_{5,\bk}\omega_{2,\bk})/\sqrt{\beta_{1,\bk}}}{{\ri}\omega_m+\varepsilon_{2,\bk}}
           \bigg)\\
\nonumber
\fl\!\!&-& \frac 1{\omega_{3,\bk}}\bigg(
                    \frac{y_{6,\bk}{+}(x_{6,\bk}{+}z_{6,\bk}\omega_{3,\bk})/\sqrt{\beta_{2,\bk}}}{{\ri}\omega_m-\varepsilon_{3,\bk}}
                +\frac{{-}y_{6,\bk}{-}(x_{6,\bk}{-}z_{6,\bk}\omega_{3,\bk})/\sqrt{\beta_{2,\bk}}}{{\ri}\omega_m+\varepsilon_{3,\bk}}
           \bigg)\\
\nonumber
\fl\!\!&-& \frac 1{\omega_{4,\bk}}\bigg(
                    \frac{y_{6,\bk}{-}(x_{6,\bk}{+}z_{6,\bk}\omega_{4,\bk})/\sqrt{\beta_{2,\bk}}}{{\ri}\omega_m-\varepsilon_{4,\bk}}
                +\frac{{-}y_{6,\bk}{+}(x_{6,\bk}{-}z_{6,\bk}\omega_{4,\bk})/\sqrt{\beta_{2,\bk}}}{{\ri}\omega_m+\varepsilon_{4,\bk}}
           \bigg)
\bigg\}\\
\eea
\bea
\nonumber
\fl G^-_{32}(\bk,\omega_m)=\frac{\eta}{4}\bigg\{\!\!&\hphantom{+}&
         \frac 1{\omega_{1,\bk}}\bigg(
                    \frac{y_{7,\bk}{+}(x_{7,\bk}{+}z_{7,\bk}\omega_{1,\bk})/\sqrt{\beta_{1,\bk}}}{{\ri}\omega_m-\varepsilon_{1,\bk}}
                +\frac{{-}y_{7,\bk}{-}(x_{7,\bk}{-}z_{7,\bk}\omega_{1,\bk})/\sqrt{\beta_{1,\bk}}}{{\ri}\omega_m+\varepsilon_{1,\bk}}
           \bigg)\\
\nonumber
\fl\!\!&+& \frac 1{\omega_{2,\bk}}\bigg(
                    \frac{y_{7,\bk}{-}(x_{7,\bk}{+}z_{7,\bk}\omega_{2,\bk})/\sqrt{\beta_{1,\bk}}}{{\ri}\omega_m-\varepsilon_{2,\bk}}
                +\frac{{-}y_{7,\bk}{+}(x_{7,\bk}{-}z_{7,\bk}\omega_{2,\bk})/\sqrt{\beta_{1,\bk}}}{{\ri}\omega_m+\varepsilon_{2,\bk}}
           \bigg)\\
\nonumber
\fl\!\!&-& \frac 1{\omega_{3,\bk}}\bigg(
                    \frac{y_{8,\bk}{+}(x_{8,\bk}{+}z_{8,\bk}\omega_{3,\bk})/\sqrt{\beta_{2,\bk}}}{{\ri}\omega_m-\varepsilon_{3,\bk}}
                +\frac{{-}y_{8,\bk}{-}(x_{8,\bk}{-}z_{8,\bk}\omega_{3,\bk})/\sqrt{\beta_{2,\bk}}}{{\ri}\omega_m+\varepsilon_{3,\bk}}
           \bigg)\\
\nonumber
\fl\!\!&-& \frac 1{\omega_{4,\bk}}\bigg(
                    \frac{y_{8,\bk}{-}(x_{8,\bk}{+}z_{8,\bk}\omega_{4,\bk})/\sqrt{\beta_{2,\bk}}}{{\ri}\omega_m-\varepsilon_{4,\bk}}
                +\frac{{-}y_{8,\bk}{+}(x_{8,\bk}{-}z_{8,\bk}\omega_{4,\bk})/\sqrt{\beta_{2,\bk}}}{{\ri}\omega_m+\varepsilon_{4,\bk}}
           \bigg)
\bigg\}\\
\fl\\
\nonumber
\fl G^-_{42}(\bk,\omega_m)=\frac{\eta}{4}\bigg\{\!\!&\hphantom{+}&
         -\frac {1}{\omega_{1,\bk}}\bigg(
                    \frac{y_{7,\bk}{+}(x_{7,\bk}{+}z_{7,\bk}\omega_{1,\bk})/\sqrt{\beta_{1,\bk}}}{{\ri}\omega_m-\varepsilon_{1,\bk}}
                +\frac{{-}y_{7,\bk}{-}(x_{7,\bk}{-}z_{7,\bk}\omega_{1,\bk})/\sqrt{\beta_{1,\bk}}}{{\ri}\omega_m+\varepsilon_{1,\bk}}
           \bigg)\\
\nonumber
\fl\!\!&-& \frac 1{\omega_{2,\bk}}\bigg(
                    \frac{y_{7,\bk}{-}(x_{7,\bk}{+}z_{7,\bk}\omega_{2,\bk})/\sqrt{\beta_{1,\bk}}}{{\ri}\omega_m-\varepsilon_{2,\bk}}
                +\frac{{-}y_{7,\bk}{+}(x_{7,\bk}{-}z_{7,\bk}\omega_{2,\bk})/\sqrt{\beta_{1,\bk}}}{{\ri}\omega_m+\varepsilon_{2,\bk}}
           \bigg)\\
\nonumber
\fl\!\!&-& \frac 1{\omega_{3,\bk}}\bigg(
                    \frac{y_{8,\bk}{+}(x_{8,\bk}{+}z_{8,\bk}\omega_{3,\bk})/\sqrt{\beta_{2,\bk}}}{{\ri}\omega_m-\varepsilon_{3,\bk}}
                +\frac{{-}y_{8,\bk}{-}(x_{8,\bk}{-}z_{8,\bk}\omega_{3,\bk})/\sqrt{\beta_{2,\bk}}}{{\ri}\omega_m+\varepsilon_{3,\bk}}
           \bigg)\\
\nonumber
\fl\!\!&-& \frac 1{\omega_{4,\bk}}\bigg(
                    \frac{y_{8,\bk}{-}(x_{8,\bk}{+}z_{8,\bk}\omega_{4,\bk})/\sqrt{\beta_{2,\bk}}}{{\ri}\omega_m-\varepsilon_{4,\bk}}
                +\frac{{-}y_{8,\bk}{+}(x_{8,\bk}{-}z_{8,\bk}\omega_{4,\bk})/\sqrt{\beta_{2,\bk}}}{{\ri}\omega_m+\varepsilon_{4,\bk}}
           \bigg)
\bigg\}\\
\eea

\section{First-order Green's functions}
\label{eq:solutionG1}
 From the whole set of the first-order Green's functions we need only the
diagonal $G^{(1)}_{ll}$ ones.
 The solution for such Green's functions can be presented via the coefficients $V_{ln}$
(defined in equation~(\ref{eq:coef_G1})) and zero-order Green's functions in the following form:
\bea
\label{eq:G11}
\fl && G^{(1)}_{11}=-{\cal Z}\left\{\:
              |G_{12}|^2V_{22} + |G_{22}|^2V_{12} + |G^-_{12}|^2V^-_{22} + |G^-_{22}|^2V^-_{12}\right.\\
\nonumber
\fl &&\hspace*{1.6cm}\left.\:
           +  |G_{32}|^2V_{42} + |G_{42}|^2V_{32} + |G^-_{32}|^2V^-_{42} + |G^-_{42}|^2V^-_{32}
                \:\right\}\\
\label{eq:G22}
\fl && G^{(1)}_{22}=-{\cal Z}\left\{\:
              |G_{12}|^2V_{12} + |G_{22}|^2V_{22} + |G^-_{12}|^2V^-_{12} + |G^-_{22}|^2V^-_{22}\right.\\
\nonumber
\fl &&\hspace*{1.6cm}\left.\:
           +  |G_{32}|^2V_{32} + |G_{42}|^2V_{42} + |G^-_{32}|^2V^-_{32} + |G^-_{42}|^2V^-_{42}
                \:\right\}\\
\label{eq:G33}
\fl && G^{(1)}_{33}=-{\cal Z}\left\{\:
              |G_{12}|^2V_{42} + |G_{22}|^2V_{32} + |G^-_{12}|^2V^-_{42} + |G^-_{22}|^2V^-_{32}\right.\\
\nonumber
\fl &&\hspace*{1.6cm}\left.\:
           +  |G_{32}|^2V_{22} + |G_{42}|^2V_{12} + |G^-_{32}|^2V^-_{22} + |G^-_{42}|^2V^-_{12}
                \:\right\}\\
\label{eq:G44}
\fl && G^{(1)}_{44}=-{\cal Z}\left\{\:
              |G_{12}|^2V_{32} + |G_{22}|^2V_{42} + |G^-_{12}|^2V^-_{32} + |G^-_{22}|^2V^-_{42}\right.\\
\nonumber
\fl &&\hspace*{1.6cm}\left.\:
           +  |G_{32}|^2V_{12} + |G_{42}|^2V_{22} + |G^-_{32}|^2V^-_{12} + |G^-_{42}|^2V^-_{22}
                \:\right\}
\eea

\section{Transverse components of  the susceptibility in the characteristic representation}
\label{trans_comp}
 We can find the transverse components of susceptibility from the
following relations
\bea
\fl
\chi^{\sigma^x\sigma^x}_{ln}&=&\frac 12\{\hphantom{-}\Re\chi^{\sigma^+\sigma^-}_{ln}+\Re\chi^{\sigma^-\sigma^-}_{ln}\},\quad
\chi^{\sigma^y\sigma^y}_{ln} = \frac 12\{\Re\chi^{\sigma^+\sigma^-}_{ln}-\Re\chi^{\sigma^-\sigma^-}_{ln}\},\\
\fl
\chi^{\sigma^x\sigma^y}_{ln}&=&\frac 12\{          - \Im\chi^{\sigma^+\sigma^-}_{ln}-\Im\chi^{\sigma^-\sigma^-}_{ln}\},\quad
\chi^{\sigma^y\sigma^x}_{ln} = \frac 12\{\Im\chi^{\sigma^+\sigma^-}_{ln}-\Im\chi^{\sigma^-\sigma^-}_{ln}\}~~,
\eea
for any sublattice $l,n=1,2,3,4$.
 The components $\chi^{\sigma^+\sigma^-}\!,$ $\chi^{\sigma^-\sigma^-}$ can be found immediately from the solution
of the zero-order Green's functions in \ref{eq:solutionG} and the definition in equation~(\ref{eq:def}) with $\alpha=+,-$:
\bea
\fl \chi^{\sigma^+\sigma^-}_{11}=\frac{1}{4{\cal Z}}\bigg\{\!\!&\hphantom{+}&
                \frac{y_{1}{+}x_{1}/\sqrt{\beta_{1}}}{\omega^2_{1}}
                +\frac{y_{1}{-}x_{1}/\sqrt{\beta_{1}}}{\omega^2_{2}}
                +\frac{y_{2}{+}x_{2}/\sqrt{\beta_{2}}}{\omega^2_{3}}
                +\frac{y_{2}{-}x_{2}/\sqrt{\beta_{2}}}{\omega^2_{4}}
\bigg\},\\
\fl \chi^{\sigma^+\sigma^-}_{12}=\frac{1}{4{\cal Z}}\bigg\{\!\!&-&
                 \frac{y_{1}{+}x_{1}/\sqrt{\beta_{1}}}{\omega^2_{1}}
                -\frac{y_{1}{-}x_{1}/\sqrt{\beta_{1}}}{\omega^2_{2}}
                +\frac{y_{2}{+}x_{2}/\sqrt{\beta_{2}}}{\omega^2_{3}}
                +\frac{y_{2}{-}x_{2}/\sqrt{\beta_{2}}}{\omega^2_{4}}
\bigg\},\\
\fl \chi^{\sigma^+\sigma^-}_{13}=\frac{1}{4{\cal Z}}\bigg\{\!\!&\hphantom{+}&
                 \frac{y_{5}{+}x_{5}/\sqrt{\beta_{1}}}{\omega^2_{1}}
                +\frac{y_{5}{-}x_{5}/\sqrt{\beta_{1}}}{\omega^2_{2}}
                +\frac{y_{6}{+}x_{6}/\sqrt{\beta_{2}}}{\omega^2_{3}}
                +\frac{y_{6}{-}x_{6}/\sqrt{\beta_{2}}}{\omega^2_{4}}
\bigg\},\\
\fl \chi^{\sigma^+\sigma^-}_{14}=\frac{1}{4{\cal Z}}\bigg\{\!\!&-&
                 \frac{y_{5}{+}x_{5}/\sqrt{\beta_{1}}}{\omega^2_{1}}
                -\frac{y_{5}{-}x_{5}/\sqrt{\beta_{1}}}{\omega^2_{2}}
                +\frac{y_{6}{+}x_{6}/\sqrt{\beta_{2}}}{\omega^2_{3}}
                +\frac{y_{6}{-}x_{6}/\sqrt{\beta_{2}}}{\omega^2_{4}}
\bigg\},\\
\fl \chi^{\sigma^-\sigma^-}_{11}=\frac{1}{4{\cal Z}}\bigg\{\!\!&\hphantom{+}&
                 \frac{y_{3}{+}x_{3}/\sqrt{\beta_{1}}}{\omega^2_{1}}
                +\frac{y_{3}{-}x_{3}/\sqrt{\beta_{1}}}{\omega^2_{2}}
                +\frac{y_{4}{+}x_{4}/\sqrt{\beta_{2}}}{\omega^2_{3}}
                +\frac{y_{4}{-}x_{4}/\sqrt{\beta_{2}}}{\omega^2_{4}}
\bigg\},\\
\fl \chi^{\sigma^-\sigma^-}_{12}=\frac{1}{4{\cal Z}}\bigg\{\!\!&-&
                 \frac{y_{3}{+}x_{3}/\sqrt{\beta_{1}}}{\omega^2_{1}}
                -\frac{y_{3}{-}x_{3}/\sqrt{\beta_{1}}}{\omega^2_{2}}
                +\frac{y_{4}{+}x_{4}/\sqrt{\beta_{2}}}{\omega^2_{3}}
                +\frac{y_{4}{-}x_{4}/\sqrt{\beta_{2}}}{\omega^2_{4}}
\bigg\},\\
\fl \chi^{\sigma^-\sigma^-}_{13}=\frac{1}{4{\cal Z}}\bigg\{\!\!&\hphantom{+}&
                 \frac{y_{7}{+}x_{7}/\sqrt{\beta_{1}}}{\omega^2_{1}}
                +\frac{y_{7}{-}x_{7}/\sqrt{\beta_{1}}}{\omega^2_{2}}
                +\frac{y_{8}{+}x_{8}/\sqrt{\beta_{2}}}{\omega^2_{3}}
                +\frac{y_{8}{-}x_{8}/\sqrt{\beta_{2}}}{\omega^2_{4}}
\bigg\},\\
\fl \chi^{\sigma^-\sigma^-}_{14}=\frac{1}{4{\cal Z}}\bigg\{\!\!&-&
                 \frac{y_{7}{+}x_{7}/\sqrt{\beta_{1}}}{\omega^2_{1}}
                -\frac{y_{7}{-}x_{7}/\sqrt{\beta_{1}}}{\omega^2_{2}}
                +\frac{y_{8}{+}x_{8}/\sqrt{\beta_{2}}}{\omega^2_{3}}
                +\frac{y_{8}{-}x_{8}/\sqrt{\beta_{2}}}{\omega^2_{4}}
\bigg\},
\eea
where all coefficients $x_l$, $y_l$ with $l=1,\dots 8$, which are taken from the \ref{eq:solutionG},
and all frequencies $\omega$, and $\beta_{1,2}$ (see equations~(\ref{eq:varepsilon},\ref{eq:beta}))
are taken in the long wavelength limit $\bk\rightarrow0$.
 Note that in above formulae we have also used the following relations between the different components
of the transverse components of susceptibility in the CR
\bea
&&\chi_{11}=\chi_{22}=\chi_{33}=\chi_{44},\qquad
  \chi_{12}=\chi_{21}=\chi_{34}=\chi_{43},\\
\nonumber
&&\chi_{13}=\chi_{31}=\chi_{42}=\chi_{24},\qquad
  \chi_{14}=\chi_{41}=\chi_{32}=\chi_{23}.
\eea


\newpage

\section*{References}

\begin{thebibliography}{10}
%
\bibitem{Lavrov}
Lavrov A N, Ando Y, Komiya S and Tsukada I 2001 {\it Phys. Rev. Lett.}, {\bf 87} 0170071
%
\bibitem{jtran05}
Tranquada J M 2005 {\it cond-mat/0508272}
%
\bibitem{Tabunshchyk}
Tabunshchyk K V and Gooding R J 2005 {\it Phys. Rev. B}, {\bf 71} 214418
%
\bibitem{Dzyaloshinskii}
Dzialoshinski I 1958 {\it J. Phys. Chem. Solids}, {\bf 4} 241
%
\bibitem{Moriya}
Moriya T 1960 {\it Phys. Rev.}, {\bf 120} 91
%
\bibitem{Katanin}
Katanin A A and Kampf A P 2002 {\it Phys. Rev. B} {\bf 66}, 100403(R)
%
\bibitem{Johnston}
Johnston David C 1997 {\it Handbook of Magnetic Materials} {\bf 10} (Elsevier, New York, 1997)
%
\bibitem{Coffey}
Coffey D, Rice T M and Zhang F C 1991 {\it Phys. Rev. B}, {\bf 44} 10112
%
\bibitem{Aharony}
Shekhtman L, Ebtin-Wohlman O and Aharony A 1993 {\it Phys. Rev. Lett.}, {\bf 71} 468
%
\bibitem{Koshibae}
Koshibae W, Ohta Y and Maekawa S 1994 {\it Phys. Rev. B}, {\bf 50} 3767
%
\bibitem{Xue}
Xue W, Grest G S, Cohen M H, Sinha S K and Soukoulis C 1988 {\it Phys. Rev. B}, {\bf 38} 6868
%
\bibitem{Liu}
Lee K H, Liu S H 1967 {\it Phys. Rev.}, {\bf 159} 390
%
\bibitem{Tyablikov}
Tyablikov S V 1959 {\it Ukrain. Mat. Zh}, {\bf 11} 287
%
\bibitem{Neto}
Silva Neto M B, Benfatto L, Juricic V and Morais Smith C 2005 {\it cond-mat/0502588}
%
\bibitem{Keimer}
Keimer B, Birgeneau R J, Cassanho A, Endoh Y, Greven M, Kastner M A and Shirane G 1993
{\it Z. Phys. B}, {\bf 91} 373
%
\bibitem{Peters}
Peters C J, Birgeneau R J, Kastner M A, Yoshizawa H, Endoh Y,
Tranquada J, Shirane G, Hidaka Y, Oda M, Suzuki M and Murakami T 1988
{\it Phys. Rev. B}, {\bf 37} 9761
\end{thebibliography}
\end{document}